\documentclass[twocolumn]{jpsj2}

\title{%
Quantum and Topological Criticalities of Lifshitz Transition in Two-Dimensional Correlated Electron Systems
}

\author{%
Youhei \textsc{Yamaji}\thanks{E-mail:yamaji@solis.t.u-tokyo.ac.jp},
Takahiro \textsc{Misawa}, and Masatoshi \textsc{Imada}  
}

\inst{
 {\it 
Department of Applied Physics, University of Tokyo, 7-3-1 Hongo, Bunkyo-ku, Tokyo, 113-8656, Japan}
}

\recdate{June 26, 2006}

\abst{
We study electron correlation effects on quantum criticalities of Lifshitz transitions at zero temperature, using the mean-field theory based on a preexisting symmetry-broken order, in two-dimensional systems.
In the presence of interactions, Lifshitz transitions may become discontinuous 
in contrast to the continuous transition in the original proposal by Lifshitz for noninteracting systems.
We focus on the quantum criticality at the endpoint of discontinuous Lifshitz transitions, which we call the {\it marginal quantum critical point}. It shows remarkable criticalities arising from its nature as a topological transition.
 At the point, for the {\it canonical ensemble}, the susceptibility of the order parameter $\chi$ is found to diverge as $\ln 1/|\delta\Delta|$ when the ``neck'' of the Fermi surface 
 collapses at the van Hove singularity. More remarkably, it diverges as $|\delta\Delta|^{-1}$ when the electron/hole pocket of the Fermi surface vanishes. Here $\delta\Delta$ is the amplitude of 
the mean field measured from the Lifshitz critical point. 
On the other hand, for the {\it grand canonical ensemble}, the discontinuous transitions appear as the electronic phase separation and the endpoint of the phase separation is the marginal quantum critical point. Especially, when a pocket of the Fermi surface vanishes, the uniform charge compressibility $\kappa$ diverges as $|\delta n|^{-1}$, instead of $\chi$, where $\delta n$ is the electron density measured from the critical point. Accordingly, 
Lifshitz transition induces large fluctuations represented by diverging $\chi$ and/or $\kappa$. Such fluctuations must be involved in physics of competing orders and influence diversity of strong correlation effects.
}

\kword{
Lifshitz transition, Fermi surface topology, metamagnetism, charge compressibility, nonanalytic expansion
}

\begin{document}
\maketitle
\section{Introduction}
Lifshitz transition is one of the typical continuous quantum phase transitions~\cite{Lifshitz60} and has been extensively studied for several decades~\cite{Makarov65,Brandt66,Chu70,Struzhkin97,Egorov82,Egorov83}. However, despite the increasing interest in quantum fluctuations and resultant phenomena, the Lifshitz transition has attracted relatively less attention as a quantum phase transition, in comparison with those of quantum critical points observed, for example in the itinerant electron magnets~\cite{Sachdev99,Hertz76,Millis93}. 
For conventional magnetic transitions, quantum criticalities emerge by reducing finite critical temperatures to zero, beyond which the transition disappears. On the other hand, as we will show in this paper, the Lifshitz criticality has features very different from that of the conventional magnetic phase transitions. For example,  because of its topological nature, the Lifshitz transition does not disappear even when parameters are controlled beyond the point of vanishing critical temperature as we will clarify later.
\par
\begin{figure}[h]
\begin{center}
\includegraphics[width=6cm]{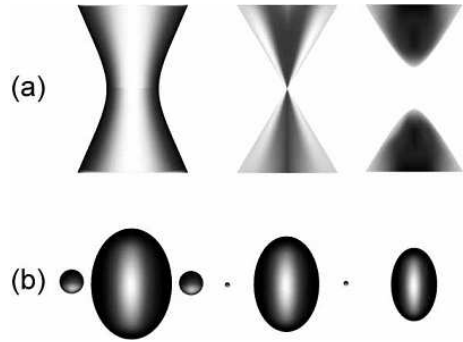}
\end{center}
\caption[]{(a) Collapse of ``neck'' of a Fermi surface.
 (b) Appearance/disappearance of a new split-off region of surface~\cite{Lifshitz60}.}
\label{Lifshitz}
\end{figure}

The Lifshitz transition occurs in noninteracting Fermion systems, which is characterized by the topological change of the Fermi surface in the Brillouin zone. Lifshitz~\cite{Lifshitz60} classified the changes of the topology of the Fermi surface in two types 
as are illustrated in Figs. \ref{Lifshitz} (a) and (b). 
The first type is the case that the ``neck'' of the Fermi surface collapses.
An appearance/disappearance of new split-off region of the Fermi surface is the other case. We call the first type of Lifshitz transition {\it neck-collapsing} type and the second one {\it pocket-vanishing} type. 
\par Because the topology of the Fermi surface is well defined only at $T=0$, 
Lifshitz critical point/line, in the original sense for noninteracting fermions, exists only at $T=0$.
Furthermore, the transition is always continuous. 
These are clearly different from the conventional quantum critical points which originate from the suppression of critical temperatures as described above. The singularities seen in the Lifshitz transition are also different from those of the conventional quantum critical points.
The singularities of the Lifshitz transitions can be seen as singular parts of 
thermodynamic potentials, $\Omega_{{\rm sing}}$. In three dimensions, the singular part is proportional to $|\zeta|^{5/2}$~\cite{Lifshitz60}. The control parameter $\zeta$ is the Fermi level, $\varepsilon_{F}$, measured from the critical point value, $\varepsilon_{F}^{c}$; $\zeta=\varepsilon_{F}-\varepsilon_{F}^{c}$. 
In two dimensions, the singular part behaves as $|\zeta|^{2}\ln 1/|\zeta|$ for the neck-collapsing type and behave as $\zeta |\zeta|$ for the pocket-vanishing type. In fact these are derived from the well-known density of states in two dimensions either with the van Hove singularity or the band edge. Namely, the singular part of the density of states $D_{{\rm sing}}$ behaves as $|\zeta|^{1/2}$ in three dimensions while in two dimensions it scales as $\ln 1/|\zeta|$ for the neck-collapsing type and scales as Heaviside functions for the pocket-vanishing type. Here Heaviside function $\theta (x)$ is defined as follows; for $x>0$, $\theta (x)$ is equal to 1 and, for $x < 0$, $\theta (x)$ is equal to 0. The singular part of the thermodynamic potential $\Omega_{{\rm sing}}$ is related with the density of states as $D_{{\rm sing}}=\partial^{2}\Omega_{{\rm sing}}/\partial \zeta^{2}$, which leads to the above singularity. The charge compressibility $\kappa$ and the uniform magnetic susceptibility $\chi$ are scaled in the same fashion as $D_{{\rm sing}}$. It has been emphasized that Lifshitz transition does not follow the conventional Ginzburg-Landau-Wilson (GLW) scheme~\cite{Wen04}, because the thermodynamic potential cannot be given by an analytic expansion of the control parameter $\zeta$
. This violation of GLW scheme is in sharp contrast with the quantum critical point of conventional magnetic transitions. These singularities are listed in Table \ref{scLT}.
\begin{table}[h]
\begin{center}
\begin{tabular}[c]{cccc}
  \hline
  &$d=3$ & $d=2$,& $d=2$,\\
  & &neck-collapsing &pocket-vanishing \\

  \hline
$\Omega_{{\rm sing}}$&$|\zeta|^{5/2}$&$|\zeta|^{2}\ln 1/|\zeta|$&$\zeta |\zeta|$ \\
$D_{{\rm sing}}$ & $|\zeta|^{1/2}$  & $\ln 1/|\zeta|$ & $\theta (\pm\zeta)$ \\
$\kappa$ & $|\zeta|^{1/2}$  & $\ln 1/|\zeta|$ & $\theta (\pm\zeta)$ \\
$\chi$ & $|\zeta|^{1/2}$ & $\ln 1/|\zeta|$ & $\theta (\pm\zeta)$\\
\hline
\end{tabular}
\end{center}
\caption[]{Singularities of conventional Lifshitz transitions. The singular parts of thermodynamic potentials $\Omega_{{\rm sing}}$ and of densities of states $D_{{\rm sing}}$ are given as functions of the Fermi level measured from critical points $\zeta$. The charge compressibility $\kappa$ and the uniform susceptibility
 $\chi$ behave as $D_{{\rm sing}}$.
 In three dimensions, all the exponents are common between neck-collapsing and pocket-vanishing cases as has already been pointed by Lifshitz~\cite{Lifshitz60}.
 }
\label{scLT}
\end{table}\par
Lifshitz considered the three-dimensional case and called the anomaly ``transitions of the 2$\frac{1}{2}$ order'' based on the terminology of Ehrenfest; the second derivatives of the electronic thermodynamic potentials have non-diverging square-root singularities and the third derivatives have diverging inverse-square-root singularities.\par
Soon after the original remarks by Lifshitz, experimental 
measurements of Lifshitz transitions were reported. The nonlinear pressure dependences of the superconducting transition temperatures $T_{c}$ were found for thallium and interpreted as the effects of Lifshitz transition~\cite{Makarov65,Brandt66}. For Li-Mg alloys, crossover effects from the Lifshitz transition were also detected by thermopower measurements~\cite{Egorov82,Egorov83}.\par
\begin{figure}[htbp]
\begin{center}
\includegraphics[width=7cm]{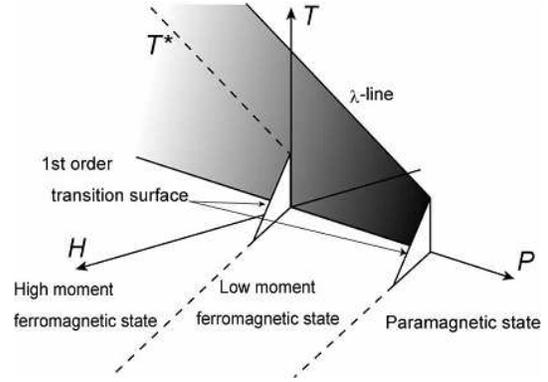}
\caption[]{Schematic phase diagram proposed for intermetallic 
 compounds ${\rm ZrZn}_{2}$ and ${\rm UGe}_{2}$~\cite{ZrZn2_dHvA}.
 $P$ and $H$ stand for applied hydrostatic pressures and external magnetic fields, respectively. The broken lines on the $P$-$H$ surface stand for the ``crossover'' lines, which separates the paramagnetic states, the low moment ferromagnetic phase and the high moment ferromagnetic phase (see text). $T^{\ast}$ means the temperature where a sudden increase of the magnetization occurs.}
\label{metamag}
\end{center}
\end{figure}

Recently, the topology of the normal state Fermi surfaces of the high-temperature superconducting cuprates (HTSC) has attracted considerable attention.
From the beginning of the studies on the cuprate superconductors, it has been well known that the Hall coefficient changes sign near the so-called optimum doping for the highest superconducting transition temperatures~\cite{Ong87,Takagi89}. This is consistent with the topological change of the Fermi surface from a hole-like to an electron-like ones. A dramatic change of metallic properties from underdoped to overdoped regions suggests possible important role of such Lifshitz transitions, although its profound effect has not fully been elucidated.\par
Among various HTSCs, the topology of the Fermi surface of bilayered cuprates ${\rm Bi_{2}Sr_{2}CaCu_{2}O_{8+\delta}}$ (Bi-2212) has been intensively investigated recently by angle-resolved photoemission spectroscopy (ARPES).
While some ARPES studies of Bi-2212 conclude the presence of large hole-like Fermi surfaces persisting for a wide range of dopings, other ARPES measurements suggest the change in the topology of the Fermi surface. Recently the bilayer splitting has been observed around the optimal doping~\cite{Chuang04}. The other groups also have observed the bilayer splitting and furthermore they have concluded that the change of the Fermi surface topology of the anti-bonding band occurs within the moderately overdoped region~\cite{Kaminski05}.\par
 Changes of the Fermi surface topology have also been proposed to be a driving mechanism of metamagnetic behaviors in itinerant electron ferromagnets ${\rm ZrZn}_{2}$ and ${\rm UGe}_{2}$~\cite{Sandeman03}. Inter-metallic compounds ${\rm ZrZn}_{2}$ and ${\rm UGe}_{2}$ have been intensively studied for decades as typical itinerant electron ferromagnets. Recently, these compounds have attracted renewed interest because of the superconductivity and multiple quantum phase transitions~\cite{Uhlarz04}. The metamagnetic behaviors and an accompanying anomalous change of the effective mass of quasiparticles are observed within the ferromagnetic phase, under applied magnetic fields. Shown in Figure \ref{metamag}, discontinuous transitions separating the phase with a relatively low magnetic moment and that with a high moment are also observed~\cite{Pfleiderer02}.
One possible mechanism of such discontinuous metamagnetic transitions is 
the topological changes of the Fermi surfaces. 
If it is true, it means that Lifshitz transitions can become discontinuous one, which contradicts the absence of finite-temperature Lifshitz transitions predicted for noninteracting systems as described above.\par
The electron correlations have been considered to play crucial roles in various  phenomena observed in the cuprates. In the itinerant electron ferromagnets, the electron correlations are also important. On the other hand, the original idea of Lifshitz transition was based on the single-particle picture. The electron correlation  plays no role in the discussion by Lifshitz. Can the electron correlations play relevant roles in the Lifshitz criticalities ? If the nature of the Lifshitz criticalities changes under the electron correlation effects, how do the criticalities change ?\par
In the present paper, 
%
%
\begin{table}[h]
\begin{center}
\begin{tabular}[c]{ccc}
  \hline
&MQCP, $d=2$,&MQCP, $d=2$,\\
&neck-collapsing&pocket-vanishing \\
\hline
$D_{{\rm sing}}$&$\ln 1/|\zeta|$
&$\theta (\pm \zeta)$\\
$\kappa$& $\ln 1/|\zeta|$& $|\zeta|^{-1}$\\
$\chi$ & $\ln 1/|\zeta|$ & $|\zeta|^{-1}$\\
\hline\hline
\end{tabular}
\end{center}
\caption[]{Singularities of Lifshitz transitions. The singular parts of densities of states $D_{{\rm sing}}$ are given as functions of the Fermi level, $\zeta$, measured from the critical points. The charge compressibility $\kappa$ and the uniform susceptibility $\chi$ behave as $D_{{\rm sing}}$ in noninteracting systems while they clearly differ at MQCP for the pocket-vanishing type. 
MQCP stands for the endpoint of discontinuous Lifshitz transitions.
The charge compressibility $\kappa$ stands for the singularity of MQCP of the grand canonical case. On the other hand, the susceptibility $\chi$ stands for the canonical case.}
\label{table_LT}
\end{table}
to understand electron correlation effects on Lifshitz transitions, we construct a phenomenological theory by assuming a mean field arising from a preexisting symmetry-broken order.
Electronic interactions play roles in this phenomenological theory through the preexisting symmetry-broken order.
We also perform mean-field calculations on a microscopic model which exhibits metamagnetic transitions. 
In this paper, we mainly consider at zero temperature, $T=0$. Finite-temperature properties will be discussed in a separate publication.\par
For conventional metamagnetic quantum critical points, the universality class has been proposed to be the overdamped, conserving Ising type~\cite{Millis02},  which corresponds to that of the clean itinerant electron ferromagnets. This basically follows the conventional GLW scheme. In contrast, we obtain a novel quantum criticality for the metamagnetic transitions driven by Lifshitz transitions, which is represented by a nonanalytic 
free-energy expansion due to the singularities of the density of states. It clearly violates the GLW scheme. For the conventional metamagnetic quantum transition, the critical exponents are theoretically expected to be the mean-field exponents of Ising model~\cite{Sachdev99,Hertz76,Millis93}. In contrast, we obtain different exponents at the endpoint of the first-order Lifshitz transition at $T=0$, which will be called the {\it marginal quantum critical point} (MQCP) in the following section,
because it appears at the margin of the quantum critical line as we will see later. 
The difference from the conventional quantum critical endpoint within the GLW scheme arises from the topological origin of the Lifshitz transitions. Because of its topological constraint, the transition does not terminate at MQCP, but a quantum critical line at zero temperature continues. Although the finite-temperature critical point of the Lifshitz transition may be described by the Ising universality class, it is deeply modified at MQCP.\par
Results at MQCP reflect electron correlation effects. The charge compressibility $\kappa$ and the uniform magnetic susceptibility $\chi$, we will obtain are summarized in Tables \ref{table_LT}. The charge compressibility $\kappa$ stands for the singularity of MQCP of the {\it grand canonical} case. On the other hand, the susceptibility $\chi$ stands for the {\it canonical} case. The charge compressibility and the uniform magnetic susceptibility are enhanced at MQCP more strongly than the noninteracting case.
\par
%
%
In Section 2, we derive a phenomenological theory to elucidate the nature of the novel Lifshitz criticality at the endpoint of the discontinuous Lifshitz transition. Nonanalytic expansions of the free energy are obtained. These expansions clearly violate the GLW scheme. In Section 3, we present numerical studies on the $t$-$t'$Hubbard model with an external magnetic field, which exhibits such novel Lifshitz criticalities. Section 4 is devoted to summary and discussion.
\section{Free-Energy Expansion}
\subsection{Mean-field theory}
\label{Mean field theory}
We assume that there are two bands $\varepsilon_{\pm}(k,\Delta)$  arising from some symmetry breaking described by order parameter $\Delta$. For example, $\Delta$ may stand for the magnetization within the ferromagnetism and each band represents the majority and minority spin subband, respectively.
The explicit form of these spin subbands is given as,
\begin{align}
\varepsilon_{\pm}(k,\Delta)=\varepsilon_{0}(k)\pm \left|\Delta_{0}+\Delta\right|,\label{dispersion}
\end{align}
where $\varepsilon_{0}(k)$ is the bare band dispersion and $\Delta_{0}$ is a real constant, which expresses a preexisting symmetry-breaking field. The above band structures are obtained from a class of Hamiltonian $\hat{H}$ such as
\begin{align}
\hat{H}&=\sum_{k}\varepsilon_{0}(k) \left(a^{\dagger}_{k}a^{\ }_{k}+b^{\dagger}_{k}b^{\ }_{k}\right)\notag\\
&+\Delta_{0}\sum_{i}\left(a^{\dagger}_{i}b^{\ }_{i}+b^{\dagger}_{i}a^{\ }_{i}\right)
+g\sum_{i}n_{ia}n_{ib},
\end{align}
where $a^{\dagger}_{k}$($a^{\ }_{k}$) and $b^{\dagger}_{k}$($b^{\ }_{k}$) represent the creation(annihilation) operators of an electron at wave number $k$ belonging to A and B subbands, respectively. For the ferromagnetic systems, these subbands stand for the up-spin and down-spin subbands. The coupling constant is given by $g$ with the number operators of $i$-th site $n_{ia}=a^{\dagger}_{i}a^{\ }_{i}$ and $n_{ib}=b^{\dagger}_{i}b^{\ }_{i}$. 
The interaction term $g\sum_{i}n_{ia}n_{ib}$ is rewritten as
\begin{align}
&g\sum_{i}n_{ia}n_{ib}\notag\\
=&\frac{g}{N_{s}}\sum_{kqp}a^{\dagger}_{k}a^{\ }_{q}b^{\dagger}_{p}b^{\ }_{k-q+p}\notag\\
=&\frac{g}{N_{s}}\sum_{kp} a^{\dagger}_{k}a^{\ }_{k}b^{\dagger}_{p}b^{\ }_{p}-\frac{g}{N_{s}}\sum_{kp} a^{\dagger}_{k}b^{\ }_{k}b^{\dagger}_{p}a^{\ }_{p}\notag\\
&+ \frac{g}{N_{s}}\sum_{kqp; q\neq k,p}a^{\dagger}_{k}a^{\ }_{q}b^{\dagger}_{p}b^{\ }_{k-q+p}.\label{wave_number}
\end{align}
Then we are allowed to introduce possible order parameters $\Delta$ , $\overline{n}_{a}$ and $\overline{n}_{b}$ defined by $\Delta\equiv gN_{s}^{-1}\sum_{k}\langle a^{\dagger}_{k}b^{\ }_{k}\rangle$, $\overline{n}_{a}\equiv N_{s}^{-1} \sum_{k}\langle a^{\dagger}_{k}a^{\ }_{k}\rangle$
 and $\overline{n}_{b}\equiv N_{s}^{-1} \sum_{k}\langle b^{\dagger}_{k}b^{\ }_{k}\rangle$, where the average $\langle \cdots \rangle$ should be self-consistently determined, afterwards. With these order parameters, eq. (\ref{wave_number}) is decoupled to
\begin{align}
&g\sum_{i}n_{ia}n_{ib}\notag\\
\simeq&\frac{g}{N_{s}}\sum_{kp} a^{\dagger}_{k}a^{\ }_{k}b^{\dagger}_{p}b^{\ }_{p}-\frac{g}{N_{s}}\sum_{kp} a^{\dagger}_{k}b^{\ }_{k}b^{\dagger}_{p}a^{\ }_{p}\notag\\
\simeq & g\sum_{k}\left( \overline{n_{b}} a^{\dagger}_{k}a^{\ }_{k}+\overline{n_{a}} b^{\dagger}_{k}b^{\ }_{k} \right) 
- gN_{s}\overline{n}_{a}\overline{n}_{b}\notag\\
 & -\sum_{k}\left(\Delta^{\ast} a^{\dagger}_{k}b^{\ }_{k} + \Delta b^{\dagger}_{k}a^{\ }_{k}\right)
+N_{s}\frac{|\Delta|^{2}}{g}.
\end{align}
In terms of the ferromagnetism, $\Delta_{0}$ is a transverse magnetic field and ${\rm Re} \Delta$ is the magnetization parallel to $\Delta_{0}$. 
On the other hand, $|\overline{n}_{a}-\overline{n}_{b}|$ and ${\rm Im} \Delta$ stand for the components of the magnetization, which are perpendicular to $\Delta_{0}$. We may take $\overline{n}_{a}=\overline{n}_{b}=n/2$ and ${\rm Im} \Delta =0$ , because the nonzero $|\overline{n}_{a}-\overline{n}_{b}|$ and ${\rm Im} \Delta$ are not favored in terms of the energy. Here $n$ is the electron density. Then, after the Bogoliubov transformation, 
\begin{align}
\alpha^{\dagger}_{k\pm}=\frac{1}{\sqrt{2}}a^{\dagger}_{k}\mp \frac{1}{\sqrt{2}}b^{\dagger}_{k},
\end{align}
we obtain the mean-field Hamiltonian $\mathcal{H}_{MF}$ as
\begin{align}
\mathcal{H}_{MF}&=\sum_{k,\eta=\pm}\left(\varepsilon_{\eta}(k,\Delta) + \frac{gn}{2}\right)\alpha^{\dagger}_{k\eta}\alpha^{\ }_{k\eta}\notag\\
&-\frac{N_{s}gn^{2}}{4}
+N_{s}\frac{\Delta^{2}}{g}.
\end{align}
The ground-state wave function for $\mathcal{H}_{MF}$ is given as
\begin{align}
\left|\Phi_{MF}\right\rangle \equiv \prod_{ \{k|\varepsilon_{+}(k,\Delta)\le \varepsilon_{F}\} }\alpha^{\dagger}_{k+}\prod_{ \{q|\varepsilon_{-}(q,\Delta)\le \varepsilon_{F}\} }\alpha^{\dagger}_{q-} \left|0 \right\rangle ,
\end{align}
where $\varepsilon_{F}$ is the Fermi level. Then, the mean field $\Delta$ is self-consistently given as 
\begin{align}
\Delta =  \frac{g}{N_{s}}\sum_{k}\left\langle a^{\dagger}_{k}b^{\ }_{k} \right\rangle = \frac{g}{N_{s}}\sum_{k}\left\langle b^{\dagger}_{k}a^{\ }_{k} \right\rangle, \label{self-consistent-1}
\end{align}
where $\left\langle \mathcal{O}\right\rangle \equiv \left\langle \Phi_{MF}\right|\mathcal{O} \left|\Phi_{MF}\right\rangle$. The electron density $n$ is given as $n=\langle \hat{n} \rangle$.
\par
The singularities of the density of states, which arise from the changes of the Fermi surface topology, are crucial in Lifshitz transitions. In two-dimensional electron systems, the density of states for the bare dispersion $\varepsilon_{0}(k)$, namely, $D_{0}(\varepsilon)$ inevitably has singularities such as a logarithmically diverging peak and a Heaviside step function, which correspond to Lifshitz transitions of the neck-collapsing type and the pocket-vanishing type, respectively.   %
%
To discuss Lifshitz transitions, we should write the free-energy density $\mathcal{E}$ in terms of the density of states as 
\begin{align}
\mathcal{E} =& \lim_{N_{s}\rightarrow \infty} \frac{\langle \mathcal{H}_{MF}\rangle}{N_{s}}\notag\\
=& \int_{-\infty}^{\varepsilon_{F}}
d\varepsilon \varepsilon \left\{D_{-}(\varepsilon,\Delta)
+D_{+}(\varepsilon,\Delta)
\right\}\notag\\
&+\frac{\Delta ^{2}}{g}+\frac{gn^{2}}{4},
\label{fedr}
\end{align}
where the densities of states for $\varepsilon_{\pm} \left(k,\Delta\right)$ are denoted by $D_{\pm}(\varepsilon,\Delta)$.
The self-consistent equation (\ref{self-consistent-1}) can be rewritten as
\begin{align}
\Delta=\frac{g}{2}\int_{-\infty}^{\varepsilon_{F}}
d\varepsilon \left\{D_{-}(\varepsilon,\Delta)
-D_{+}(\varepsilon,\Delta)
\right\}.\label{SCE1}
\end{align}
The electron density is also rewritten as
\begin{align}
n=\int_{-\infty}^{\varepsilon_{F}}
d\varepsilon \left\{D_{-}(\varepsilon,\Delta)
+D_{+}(\varepsilon,\Delta)
\right\}.\label{eldd}
\end{align}
In the following derivations, we assume these band structures (\ref{dispersion}) and the self-consistent equation (\ref{SCE1}) for simplicity.
However, it captures the essential physics of Lifshitz transitions even for other band structures.\par
We can make the Fermi level $\varepsilon_{F}$ approach these singularities of the densities of states at $\varepsilon_{F}=\varepsilon_{F}^{L}$ by changing the amplitude of the mean field $\Delta$. When the Fermi surface topology changes, both the stable and metastable solutions of the self-consistent equation (\ref{self-consistent-1}), for a coupling constant $g=g_{L}$, are denoted by $\Delta_{L}$.
The electron density, $n$, for given $\Delta_{L}$ and $g_{L}$ is denoted by $n_{L}$.\par
For small $\delta\Delta\equiv (\Delta -\Delta_{L})$, the free-energy density $\mathcal{E}$ may be expanded in terms of $\delta\Delta$, but with different coefficients for positive $\delta\Delta$ and negative $\delta\Delta$ in general as $\mathcal{E}_{p}=a\delta\Delta+b_{p}\delta\Delta^{2}+c_{p}\delta\Delta^{3}+\mathcal{O}(\delta\Delta^{4})$ for $\delta\Delta>0$ and $\mathcal{E}_{m}=a\delta\Delta+b_{m}\delta\Delta^{2}+c_{m}\delta\Delta^{3}+\mathcal{O}(\delta\Delta^{4})$ for $\delta\Delta <0$.
Such different coefficients appear because of the nonanalyticity of the density of states as we will clarify later.
We take the same $a$ for the both sides $\delta\Delta\lessgtr 0$, since it will turn out to be the same.
For the {\it canonical} ensemble, the electron density is fixed at $n=n_{L}$. For the {\it grand canonical} ensemble, the free-energy density $\widetilde{\mathcal{E}} \equiv \mathcal{E}-\frac{\partial \mathcal{E}}{\partial \delta n}\delta n$ is similarly expanded with respect to $\delta n \equiv n-n_{L}$. 
The expansion by $\delta n$ gives the stability condition for the electronic phase separation in the {\it grand canonical} ensemble.
\subsection{Canonical ensemble}
\subsubsection{Neck-collapsing case}
\label{nc_2_1}
\begin{figure}[h]
\begin{center}
\includegraphics[width=8cm]{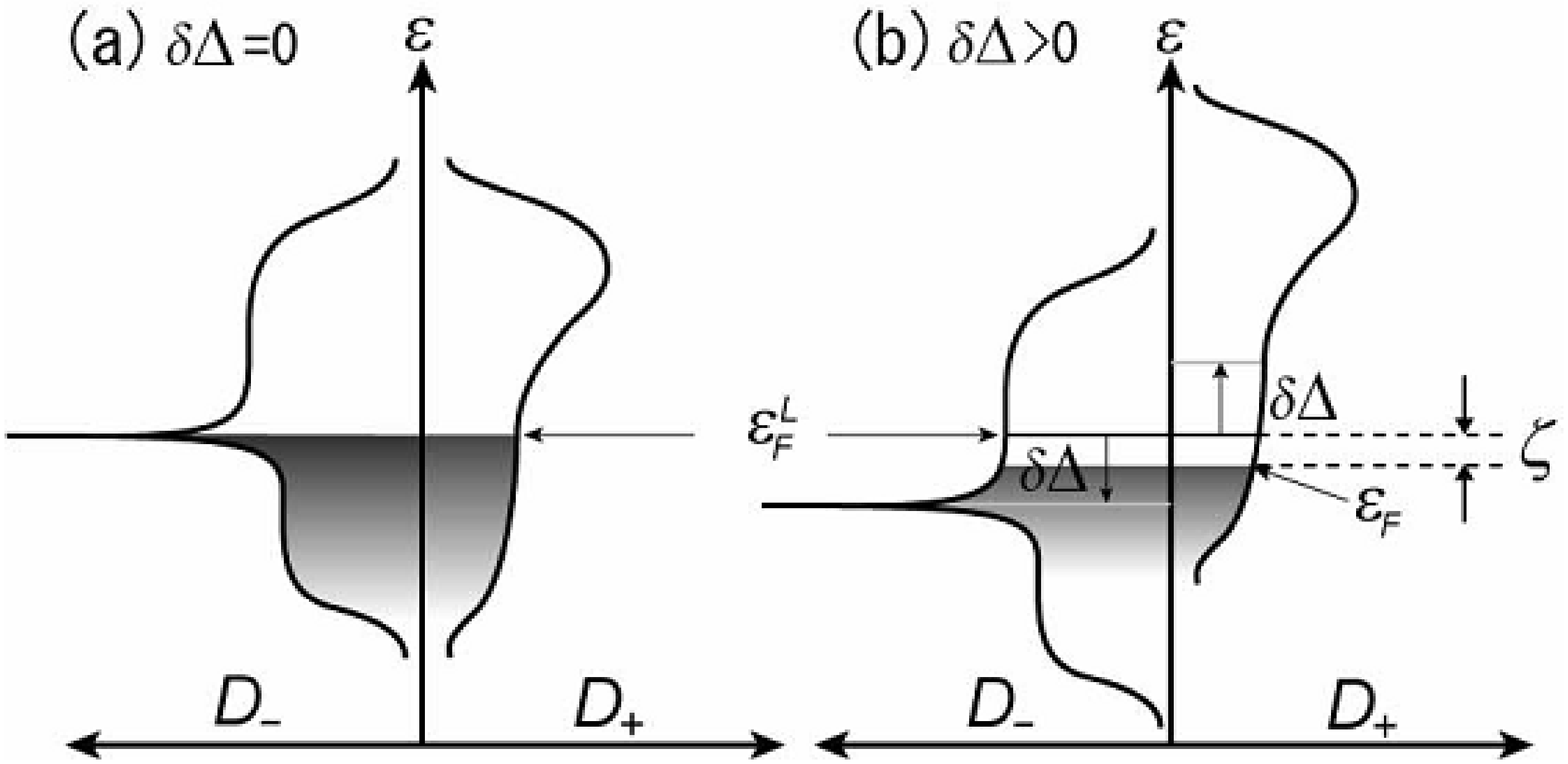}
\end{center}
\caption[]{(a)Schematic illustration of density of states $D_{\pm}(\varepsilon,\Delta)$, for $\delta\Delta =0$. (b) Schematic density of states $D_{\pm}(\varepsilon,\Delta)$, for $\delta\Delta >0$. Shaded areas are occupied states and the Fermi level shift, with an amount $\zeta = \varepsilon_{F}-\varepsilon_{F}^{L}$, is induced by the shifts of dispersions $\varepsilon_{\pm}$, namely, $\delta\Delta$.  
}
\label{D_N_C}
\end{figure}

The expansions of the free energy $\mathcal{E}$ depend on the shape of the densities of states $D_{\pm}(\varepsilon,\Delta)$. Especially, for small deviation of the mean field $\delta\Delta$, the dispersion $\varepsilon_{+}$ shifts upwards with an amount $\delta\Delta$ and $\varepsilon_{-}$ shifts downwards with an amount $-\delta\Delta$. Then the Fermi level $\varepsilon_{F}$ also shifts to keep the density $n$. Below, the amount of the Fermi level shift is denoted by $\zeta \equiv \varepsilon_{F}-\varepsilon_{F}^{L}$, as is illustrated in Fig. \ref{D_N_C}. 
Expansions of $D_{+}(\varepsilon,\Delta)$ and $D_{-}(\varepsilon,\Delta)$ around the value of Fermi level $\varepsilon_{F}^{L}$ and the mean field $\Delta_{L}$ determine the coefficients of free-energy expansion. The expansions of $D_{+}(\varepsilon,\Delta)$ and $D_{-}(\varepsilon,\Delta)$ are given as
\begin{align}
D_{+}(\varepsilon,\Delta_{L}+\delta\Delta)\simeq& d_{L+}^{(0)}+d_{L+}^{(1)}(\varepsilon - \varepsilon_{F}^{L}-\delta\Delta)\notag\\
&+\mathcal{O}\left\{(\varepsilon - \varepsilon_{F}^{L}-\delta\Delta)^{2}\right\},\label{D_+d}\\
D_{-}(\varepsilon,\Delta_{L}+\delta\Delta)\simeq& d_{L-}^{(0)} + d_{L-}^{(1)}(\varepsilon - \varepsilon_{F}^{L}+\delta\Delta)\notag\\
&+\rho_{{\rm log}}\ln \frac{1}{\left|\varepsilon - \varepsilon_{F}^{L}+\delta\Delta\right|}\notag\\
&+\mathcal{O}\left\{(\varepsilon - \varepsilon_{F}^{L}+\delta\Delta)^{2}\right\},\label{D_-d}
\end{align}
where we note that $d_{L+}^{(0)}$ and $d_{L+}^{(0)}$ depend on $\Delta_{L}$ while $d_{L-}^{(0)}$, $d_{L-}^{(1)}$ and $\rho_{{\rm log}}$ do not for the rigid band dispersions (\ref{dispersion}). Here we have used the fact that $D_{+}$ is analytic around $D_{+}(\varepsilon_{F}^{L},\Delta_{L})\equiv d_{L+}^{(0)}$ and can be expanded by the Taylor series with the first-order coefficient $d_{L+}^{(1)}$. On the other hand, $D_{-}$ has the van Hove singularity with a logarithmic singularity at $D_{-}(\varepsilon_{F}^{L},\Delta_{L})$. The analytic part of $D_{-}$ is expanded with the coefficients $d_{L-}^{(0)}$ and $d_{L-}^{(1)}$.
For $D_{-}(\varepsilon,\Delta_{L})$, the location of the logarithmically diverging peak is fixed as the origin of the expansion. On the other hand, because of the constraint of the fixed electron density, the coefficient of the expansion (\ref{D_+d}) depends on the location of the Fermi level $\varepsilon_{F}^{L}$ and accordingly on $\Delta_{L}$.\par
When the electron density is fixed at $n=n_{L}$, this constraint determines $\zeta = \varepsilon_{F}-\varepsilon_{F}^{L}$. By substituting the expansions of the density of states (\ref{D_+d}) and (\ref{D_-d}) into eq. (\ref{eldd}), we obtain the relation between $\zeta$ and $\delta\Delta$ as
\begin{align}
\zeta \sim -\delta \Delta \left(1- \frac{2d_{L+}^{(0)}}{\rho_{{\rm log}}}\frac{1}{\ln \frac{1}{\left|\delta\Delta\right|}}\right).\label{delta_mu}
\end{align}
This relation is derived in Appendix A (see eq.(\ref{delta_mu_s})).
From eq. (\ref{fedr}),
the expansion of $\mathcal{E}$ around the Lifshitz critical point is obtained as
\begin{align}
\mathcal{E}_{\eta}=& {\rm const.}\ +
a\delta \Delta
+b_{\eta}\delta \Delta^{2}\notag\\
&+c_{\eta}\frac{\delta \Delta^{2}}{\ln \frac{1}{\left|\delta \Delta\right|}}+{\rm (higher\ order\ terms)}
\label{free_energy_nc}
\end{align}
where $\eta=p,m$ and the coefficients $a$, $b_{\eta}$ and $c_{\eta}$ are given as 
\begin{align}
&a=-\frac{2\Delta_{L}}{g^{2}_{L}}\delta g,\\
&b_{p}=b_{m}=\frac{1}{g_{L}}-2d_{L+}^{(0)}+\mathcal{O}(\delta g),\label{b_p_b_m}\\
&c_{p}=c_{m}=\frac{{2d_{L+}^{(0)}}^{2}}{\rho_{{\rm log}}}.
\end{align}
The higher order terms of the order $\mathcal{O}(\delta g)$ in $b_{p}$ and $b_{m}$ do not play roles in the later discussion and can be ignored.
The derivation of eq. (\ref{free_energy_nc}) is also given in Appendix A.
The coupling constant $g$, which leads to $\Delta_{L}$ is denoted by $g_{L}$ and $\delta g$ stands for the coupling constant measured from $g_{L}$;$\delta g \equiv g-g_{L}$.
A striking feature of this expansion is that only $d_{L+}^{(0)}$ and $\rho_{{\rm log}}$ determine the expansion (\ref{free_energy_nc}). The term proportional to $\delta\Delta^{2}/\ln 1/|\delta\Delta|$ reflects the logarithmic divergence of the density of states (\ref{D_-d})~\cite{note}. 
\subsubsection{Pocket-vanishing case}
\label{pv_2_2}
\begin{figure}[h]
\begin{center}
\includegraphics[width=8cm]{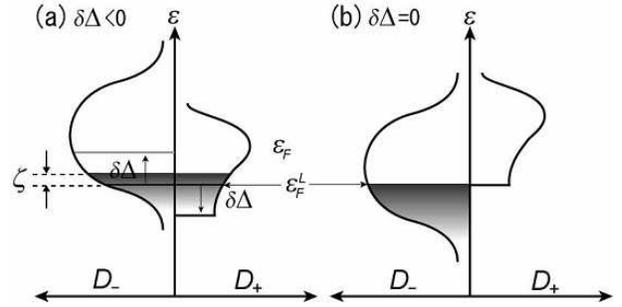}
\end{center}
\caption[]{(a)Schematic illustration of densities of states $D_{\pm}(\varepsilon,\Delta)$, for $\delta\Delta <0$. (b) Schematic densities of states $D_{\pm}(\varepsilon,\Delta)$, for $\delta\Delta =0$. Shaded areas are occupied states and the Fermi level shift, with an amount $\zeta = \varepsilon_{F}-\varepsilon_{F}^{L}$, is induced by the shifts of dispersions $\varepsilon_{\pm}$, namely, $\delta\Delta$.
}
\label{D_P_V}
\end{figure}

Instead of (\ref{D_+d}) and (\ref{D_-d}),
the expansions of $D_{\pm}(\varepsilon,\Delta_{L}+\delta\Delta)$ for the pocket-vanishing type of Lifshitz transitions are given as,
\begin{align}
D_{+}(\varepsilon,\Delta_{L}+\delta\Delta)&=&\theta (\varepsilon-\varepsilon^{L}_{F}-\delta \Delta)\left[d_{L+}^{(0)}\right.
\notag\\
&&+d_{L+}^{(1)}\left(\varepsilon-\varepsilon^{L}_{F}-\delta\Delta
\right)\notag\\
&&+
\left.
\mathcal{O}\left(\varepsilon-\varepsilon^{L}_{F}-\delta\Delta\right)^{3}
\right],\label{D_+d_pvs}\\
D_{-}(\varepsilon,\Delta_{L}+\delta\Delta)&=&d_{L-}^{(0)}+
d_{L-}^{(1)}\left(\varepsilon-\varepsilon^{L}_{F}+\delta\Delta\right)
\notag\\
&&+\mathcal{O}\left(\varepsilon-\varepsilon^{L}_{F}+\delta\Delta\right)^{3}
.\label{D_-d_pvs}
\end{align}
The shifts of these densities of states induced by $\delta\Delta$ are illustrated in Fig. \ref{D_P_V}.
Then, the condition of the fixed electron density $n=n_{L}$ gives the following relation 
\begin{align}
\zeta =& \theta (-\delta\Delta)\notag\\
\times&\left\{-\frac{d_{L-}^{(0)}-d_{L+}^{(0)}}{d_{L-}^{(0)}+d_{L+}^{(0)}}\delta\Delta +2\frac{d_{L-}^{(1)}{d_{L+}^{(0)}}^{2}-d_{L+}^{(1)}{d_{L-}^{(0)}}^{2}}{\left(d_{L-}^{(0)}+d_{L+}^{(0)}\right)^{3}}\delta\Delta^{2}+\mathcal{O}(\delta\Delta^{3})
\right\}\notag\\
&+\theta (\delta\Delta)\left(-\delta\Delta\right)
 .\label{dmu_pv}
\end{align}
Equation (\ref{dmu_pv}) is derived in Appendix A (see eq. (\ref{delmu_pv}) and below).
The free-energy expansion reflects the singular behavior of the density of states due to $\theta (-\delta \Delta)$ and is derived in Appendix A as
\begin{align}
\mathcal{E}_{\eta}=&{\rm const.}
+a\delta\Delta+b_{\eta}\delta\Delta^{2}\notag\\
&+c_{\eta}\delta\Delta^{3}+\mathcal{O}\left(\delta\Delta^{4}\right),
\label{free_energy_pv}
\end{align}
where $\eta=p,m$ and the coefficients are defined as
\begin{align}
a=&-(n_{-L}-n_{+L})+\frac{2\Delta_{L}}{g}=-\frac{2\Delta_{L}}{g^{2}_{L}}\delta g,\\
b_{m}=&\frac{1}{g_{L}}-\frac{2d_{L-}^{(0)}d_{L+}^{(0)}}{d_{L-}^{(0)}+d_{L+}^{(0)}}+\mathcal{O}(\delta g),\label{b_m_pv}\\
b_{p}=&\frac{1}{g_{L}}+\mathcal{O}(\delta g),\label{b_p_pv}\\
 c_{m}=&\frac{4}{3}\frac{d_{L+}^{(1)}{d_{L-}^{(0)}}^{3}-d_{L-}^{(1)}{d_{L+}^{(0)}}^{3}}{\left(d_{L-}^{(0)}+d_{L+}^{(0)}\right)^{3}},\\
 c_{p}=&0.
\end{align}
Similarly to eq. (\ref{b_p_b_m}) we ignore the higher order terms of $\mathcal{O}(\delta g)$ in eqs. (\ref{b_m_pv}) and (\ref{b_p_pv}) in the later discussions.
If $c_{m}<0$, higher order terms are needed for the stability of the expansion. 
\subsubsection{Phase transition}
\begin{figure}[h]
\begin{center}
\includegraphics[width=6cm]{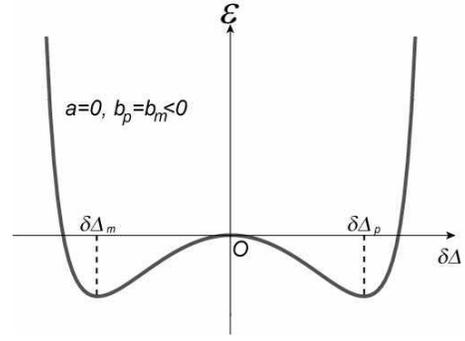}
\end{center}
\caption[]{Free-energy density $\mathcal{E}$ as a function of $\delta\Delta$, for the neck-collapsing case. For $b_{p}=b_{m}<0$, $\mathcal{E}$ has a double well structure. The value of $\delta\Delta$, namely $\delta\Delta_{p}>0$ and $\delta\Delta_{m}<0$ give the double minimum of $\mathcal{E}$.
}
\label{first_order_del}
\end{figure}

\begin{figure}[h]
\begin{center}
\includegraphics[width=8cm]{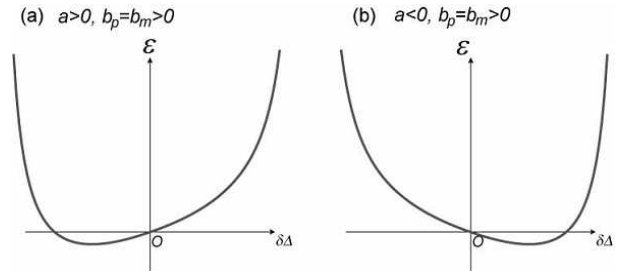}
\end{center}
\caption[]{(a) and (b): Free-energy density $\mathcal{E}$ as a function of $\delta\Delta$, for neck-collapsing case. For $b_{p}=b_{m}>0$, $\mathcal{E}$ has a single minimum at a value of $\delta\Delta$ depending on $a$. For $a>0 (a<0)$, $\mathcal{E}$ has a minimum at $\delta\Delta<0 (\delta\Delta>0)$, as is illustrated in (a) ((b)).
}
\label{continuous_del}
\end{figure}

\begin{figure}[h]
\begin{center}
\includegraphics[width=6cm]{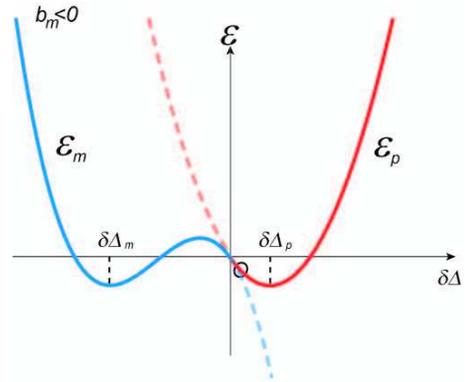}
\end{center}
\caption[]{Free-energy density $\mathcal{E}$ as a function of $\delta\Delta$ for pocket-vanishing case. The red solid curve stands for $\mathcal{E}_{p}$ and the blue solid curve stands for $\mathcal{E}_{m}$. For $b_{m}<0$, $\mathcal{E}$ has in total a double minimum at $\delta\Delta_{p}>0$ and $\delta\Delta_{m}<0$.
}
\label{first_pv_del}
\end{figure}

\begin{figure}[h]
\begin{center}
\includegraphics[width=8cm]{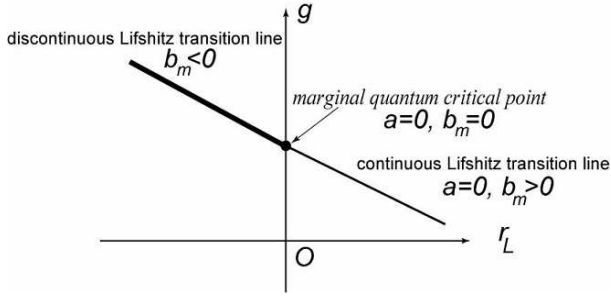}
\end{center}
\caption[]{Schematic phase diagram in plane of $r_{L}$ and $g$. MQCP is 
the point where the conditions $r_{L}=0$ and $a=0$ are satisfied. The thin solid line 
represents 
the continuous transition line and the thick solid line represents 
the discontinuous one.
}
\label{marginal}
\end{figure}

The free-energy expansions (\ref{free_energy_nc}) and (\ref{free_energy_pv}) 
have a nonanalytic form which violates the GLW scheme.
These expansions can describe continuous and first-order transitions depending on the sign of the quadratic coefficient $b_{m}$.
For the neck-collapsing transition described by eq. (\ref{free_energy_nc}), where $b_{p}=b_{m}$, the instability occurs for $b_{p}=b_{m}<0$. This leads to the first-order transition between $\delta\Delta>0$ and $\delta\Delta<0$ as we see in Fig. \ref{first_order_del}. On the other hand, the continuous transition takes place at $a=0$ when $b_{p}=b_{m}>0$ as we see in Fig. \ref{continuous_del}. 
For the pocket-vanishing transition described by eq. (\ref{free_energy_pv}), where $b_{p}> b_{m}$, the instability starts first when $b_{m}$ becomes negative as is illustrated in Fig. \ref{first_pv_del}. 
\begin{figure}[h]
\begin{center}
\includegraphics[width=8cm]{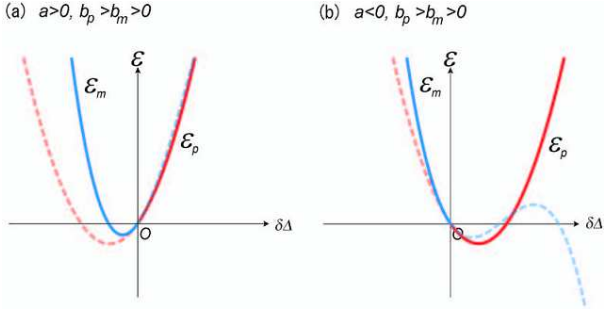}
\end{center}
\caption[]{Free-energy density $\mathcal{E}$ as a function of $\delta\Delta$ for pocket-vanishing case. The red solid curve stands for $\mathcal{E}_{p}$ and the blue solid curve stands for $\mathcal{E}_{m}$. For $b_{p}>b_{m}>0$, $\mathcal{E}$ has a single minimum at a value of $\delta\Delta$ depending on $a$. For $a>0 (a<0)$, $\mathcal{E}$ has a minimum at $\delta\Delta<0 (\delta\Delta>0)$, as is illustrated in (a) ((b)).
}
\label{continuous_pv_del}
\end{figure}
This leads to the first-order transition. The continuous transition occurs at $a=0$ if $b_{p}>b_{m}>0$ as is illustrated in Fig. \ref{continuous_pv_del}. Therefore, for both cases, the first-order transition can appear only when $b_{m}<0$. Then, it is useful to introduce a dimensionless factor $r_{L}$ called {\it Lifshitz factor} defined as
\begin{eqnarray}
r_{L}=\left\{
\begin{array}{ll}
1-2g_{L}d_{L+}^{(0)}&({\rm neck\verb|-| collapsing})\\
\displaystyle 1-2g_{L}\frac{d_{L+}^{(0)}d_{L-}^{(0)}}{d_{L+}^{(0)}+d_{L-}^{(0)}}&({\rm pocket\verb|-| vanishing})
\label{Lifhitz_canonical}
\end{array}
\right. ,
\end{eqnarray}
which is proportional to $b_{m}$.
The condition $r_{L}=0$ is analogous to the Stoner condition for the ferromagnetism. The first-order transition boundary located in the region $r_{L}<0$ and the continuous transition line located at $a=0$, $r_{L}>0$ meet at the point given by $a=r_{L}=0$ at $T=0$ as shown in Fig. \ref{marginal}. This is nothing but the {\it marginal quantum critical point} (MQCP). It indeed appears at the margin of the quantum critical line~\cite{Imada04}.
The appearance of the continuous Lifshitz line beyond MQCP at $T=0$ is ascribed to the fact that the transition is a topological one and can not terminate.
\subsubsection{Marginal quantum criticality}
The critical exponents for MQCP can be obtained from the expansions (\ref{free_energy_nc}) and (\ref{free_energy_pv}).
\paragraph{critical exponent; $\beta$}
At the first-order transition, levels of the two minima of $\mathcal{E}$ are required to be equal. The jump of the mean field $\Delta$ at the first-order transition is given from eq. (\ref{free_energy_nc}) or eq. (\ref{free_energy_pv}) as the difference between the values of $\delta\Delta$ giving the two minima of the free-energy density, namely, $\delta\Delta_{p}>0$ and $\delta\Delta_{m}<0$.\par
In the case of the neck collapsing, this requirement is automatically satisfied for $a=0$ because $\mathcal{E}$ is the even function of $\delta\Delta$. Then, the first-order transition boundary is defined by $a=0$ and $r_{L}<0$. 
For the neck-collapsing case, the jump is given from eq. (\ref{free_energy_nc}) as
\begin{eqnarray}
\delta\Delta_{p}-\delta\Delta_{m}\simeq 2\exp \left[-\frac{2{d_{L+}^{(0)}}^{2}g_{L}}{\rho_{{\rm log}}|r_{L}|}\right].\label{esse_sing}
\end{eqnarray}
When the exponent $\beta$ is defined as $\delta\Delta_{p}-\delta\Delta_{m}\propto |r_{L}|^{\beta}$, it indicates $\beta \longrightarrow \infty$ (see Appendix B). On the other hand, for the pocket-vanishing case, 
we obtain from eq. (\ref{free_energy_pv}) that
\begin{eqnarray}
a=\frac{|r_{L}|^{2}}{3c_{m}}+\mathcal{O}(r_{L}^{3})
\end{eqnarray}
and $r_{L}<0$ give the condition for the two minima $\mathcal{E}_{p}$ and $\mathcal{E}_{m}$ being equal as Fig. \ref{first_pv_del}, thus determine the first-order transition boundary.

The jump for the pocket-vanishing case is given as
\begin{eqnarray}
\delta\Delta_{p}-\delta\Delta_{m}=\frac{|r_{L}|}{3c_{m}}+\frac{g_{L}|r_{L}|^{2}}{6c_{m}}+\mathcal{O}(r_{L}^{3}),\label{beta_linear}
\end{eqnarray}
which shows $\beta =1$.
\paragraph{critical exponent; $\delta$}
At the endpoint of the first-order transition boundary defined by $r_{L}=0$,
$\delta g$-dependence of $\delta\Delta$ determines the critical exponent $\delta$ from the scaling $\delta\Delta \propto |\delta g|^{1/\delta}$. 
This relation is obtained from $\partial \mathcal{E}_{p}/\partial \delta\Delta =0$ and/or $\partial \mathcal{E}_{m}/\partial \delta\Delta =0$ at $r_{L}=0$ in eq. (\ref{free_energy_nc}).
For the neck-collapsing case, $\delta\Delta$ behaves as
\begin{eqnarray}
\delta\Delta\simeq \frac{\rho_{{\rm log}}\Delta_{L}}{2{d_{L+}^{(0)}}^{2}g_{L}^{2}}\delta g \ln \frac{1}{|\delta g|},\label{delta_log}
\end{eqnarray}
which indicates $\delta =1+0$ (see Appendix B).
On the other hand, for the pocket-vanishing case, $\delta\Delta$ is given similarly from eq. (\ref{free_energy_pv}) as
\begin{eqnarray}
|\delta\Delta | = \left\{
\begin{array}{ll} 
\sqrt[]{\frac{2\Delta_{L}}{3c_{m}g_{L}^{2}}}|\delta g|^{1/2}& (\delta g <0)\\
\frac{\Delta_{L}}{g_{L}}\delta g & (\delta g >0)
\end{array}
\right.
,\label{delta_sq}
\end{eqnarray}
which shows $\delta =2$ for $\delta g<0$ and $\delta =1$ for $\delta g>0$.
\paragraph{critical exponent; $\gamma$}
Around the endpoint $r_{L}= 0$,
the uniform susceptibility $\chi \equiv \left(\partial^{2} \mathcal{E}/\partial \delta\Delta^{2}\right)^{-1}$ behaves as
\begin{eqnarray}
\chi = g_{L} r_{L}^{-1},
\end{eqnarray}
for both of the neck-collapsing and the pocket-vanishing cases. The relation $\chi \propto r_{L}^{-\gamma}$ with $\gamma =1$ is satisfied.
However, for the the pocket-vanishing case, in the side of $\delta\Delta >0$, the uniform susceptibility $\chi$ becomes a constant, $g_{L}$, resulting in $\gamma =0$.\par
Next we consider the scaling of $\chi$ as a function of $\delta\Delta$, strictly at $r_{L}=0$. Near MQCP of the canonical ensemble, the susceptibility $\chi$ scales, as $\delta\Delta$ approaches 0, as
\begin{eqnarray}
\chi=\left\{
\begin{array}{ll}
\displaystyle \frac{1}{2c_{m}}\ln \frac{1}{|\delta\Delta|}&{\rm (neck\verb|-| collapsing)}\\
\displaystyle\frac{1}{6c_{m}|\delta\Delta|}&{\rm (pocket\verb|-| vanishing},\delta g<0)\\
\end{array}
\right. .\label{chi_D}
\end{eqnarray}

The critical exponents for both of the neck-collapsing and the pocket-vanishing cases are shown in Table \ref{critical_exponent}. Each set of exponents satisfy a scaling relation $\gamma=\beta (\delta -1)$ (see Appendix A). We note that the universality class for the pocket-vanishing type is the same as that of the metal-insulator transition~\cite{Imada04,Misawa06}.

\begin{table}[h]
\begin{center}
\begin{tabular}{cccc}
\hline
 &NC&PV  &PV  \\
 &&$(\delta g <0)$&$(\delta g>0)$\\
\hline
$\beta$&$\infty$ & 1 & 1\\
$\gamma$& 1 & 1 & 0\\
$\delta$&$1+0$ & 2 & 1\\
\hline
\end{tabular}
\end{center}
\caption[]{Critical exponents of MQCP.
	NC and PV stand for the neck-collapsing and pocket-vanishing types, respectively.
}
\label{critical_exponent}
\end{table}
\subsection{Grand canonical ensemble}
\label{grand_canonical}
For the grand canonical case, the instability towards the electronic phase separation exists. The endpoint of the electronic phase separation is nothing but MQCP in the grand canonical ensemble.
We clarify the instability by calculating the free-energy expansion with respect to $\delta n$, namely $\widetilde{\mathcal{E}}$.
For the coupling constant fixed at $g=g_{L}$, we introduce an amount of the chemical potential shift, $\delta\mu$ as a control parameter as
\begin{align}
\widetilde{\mathcal{E}}=\mathcal{E}-\frac{\partial \mathcal{E}}{\partial \delta n}-\delta\mu \delta n.
\end{align}
Then the uniform charge compressibility $\kappa$, which is given as
\begin{align}
\kappa = \left.\left(\frac{\partial^{2} \widetilde{\mathcal{E}}}{\partial \delta n^{2}}\right)\right|_{\delta n=0}^{-1},
\end{align}
exhibits a significant feature at MQCP as shown below.
\subsubsection{Neck-collapsing case}
From eq. (\ref{deln_delmu_ncnc_1}) in Appendix C, the electron density (\ref{deldel_n}) gives the relation between $\zeta$ and $\delta n$ as
\begin{align}
\zeta \simeq -\frac{g_{L}\delta n}{2r_{L}}
\left(1-\frac{2s^{2}}{g_{L}\rho_{{\rm log}}r_{L}}
\frac{1}{\ln \frac{1}{|\delta n|}}\right),\label{zeta_zeta}
\end{align}
where we assume $r_{L}\neq 0$.
Around MQCP of the grand canonical ensemble, the free-energy density $\widetilde{\mathcal{E}}$ can be expanded with respect to small $\delta n$ as is mentioned in \S \ref{Mean field theory}. The free-energy expansion of $\widetilde{\mathcal{E}}$ is given as 
\begin{eqnarray}
\widetilde{\mathcal{E}_{\eta}}=&\mathcal{E}_{\eta}-\frac{g_{L}n_{L}}{2}\delta n -\delta\mu\delta n\notag\\
\simeq& {\rm const.}+\widetilde{a}\delta n
+\widetilde{b_{\eta}}\delta n^{2}+\widetilde{c_{\eta}}\frac{\delta n^{2}}{\ln \frac{1}{|\delta n|}},\label{a_sharp_free_energy}
\end{eqnarray} 
where $\eta =p,m$ and the coefficients are defined as
\begin{align}
&\widetilde{a}=-\delta\mu,\\
&\widetilde{b_{m}}=\widetilde{b_{p}}=-\frac{g_{L}}{4}\left(\frac{1-r_{L}}{r_{L}}\right),\\
&\widetilde{c_{m}}=\widetilde{c_{p}}=-\left(d_{L-}^{(0)}+d_{L+}^{(0)}+\rho_{{\rm log}}-g_{L}^{-1}\right)\frac{sg_{L}}{2\rho_{{\rm log}}r_{L}^{2}}.
\end{align}
The derivation of eq. (\ref{a_sharp_free_energy}) is given in Appendix C (see eq. (\ref{sharp_free_energy})).

\par
\subsubsection{Pocket-vanishing case}
From eq. (\ref{deltamu_deltan}) in Appendix C, the electron density (\ref{deldel_n}) gives the relation between $\zeta$ and $\delta n$ as
\begin{align}
\zeta \sim \frac{\frac{1}{g_{L}}-\frac{d_{L-}^{(0)}+{d_{L+}^{(0)}}}{2}}{d_{L-}^{(0)}+{d_{L+}^{(0)}}}\frac{g_{L}}{r_{L}}\delta n .
\end{align}
Around MQCP of the grand canonical ensemble, the free-energy density $\widetilde{\mathcal{E}}$ can be expanded with respect to small $\delta n$ as is mentioned in \S \ref{Mean field theory}. For the sake of unified treatment, we introduce a modified form of $\delta n$ as $\widetilde{\delta n}$ defined by
\begin{align}
\widetilde{\delta n}= \delta n \cdot {\rm sign}(d_{L-}^{(0)}-d_{L+}^{(0)}),\label{widetilde_n}
\end{align}
to make the pocket Fermi surface emerge always for $\widetilde{\delta n} <0$.
 The free-energy densities $\widetilde{\mathcal{E}}_{p}$ for the positive $\widetilde{\delta n}$ and $\widetilde{\mathcal{E}}_{m}$ for the negative $\widetilde{\delta n}$ may be separately expanded with respect to $\widetilde{\delta n}$ as
\begin{align}
\widetilde{\mathcal{E}}_{\eta}=&\mathcal{E}_{\eta}-\frac{g_{L}n_{L}}{2}\delta n-\delta\mu\delta n\notag\\
\simeq&\frac{gn_{L}^{2}}{4}+\widetilde{a}\widetilde{\delta n }
+\widetilde{b_{\eta}}\widetilde{\delta n}^{2}+\widetilde{c_{\eta}}\widetilde{\delta n}^{3},
\label{expansion_dn_3}
\end{align}
where $\eta=p,m$ and the coefficients are defined as
\begin{align}
&\widetilde{a}=-\widetilde{\delta\mu},\notag\\
&\widetilde{b_{m}}=\frac{1}{2}\frac{g_{L}}{d_{L-}^{(0)}+{d_{L+}^{(0)}}}\frac{\frac{1}{g_{L}^{2}}-d_{L-}^{(0)}{d_{L+}^{(0)}}}{\frac{1}{g_{L}}-\frac{2d_{L-}^{(0)}{d_{L+}^{(0)}}}{d_{L-}^{(0)}+{d_{L+}^{(0)}}}},\\
&\widetilde{b_{p}}=\frac{1}{2}\frac{1}{d_{L-}^{(0)}},\\
&  \widetilde{c_{m}}=\frac{1}{6}\frac{1}{(d_{L-}^{(0)}+{d_{L+}^{(0)}})^{3}}\frac{1}{r^{3}}\notag\\
&\times
\left\{
\left( 2{r^{+}}^{3}+3r^{-}{r^{+}}^{2}
\right)d_{L-}^{(1)}
+\left( 2{r^{-}}^{3}+3r^{+}{r^{-}}^{2}
\right)d_{L+}^{(1)}
\right\},\notag\\
& \widetilde{c_{p}}=\frac{5}{6}\frac{1}{{d_{L-}^{(0)}}^{3}}\left(1-\frac{3}{5}g_{L}d_{L-}^{(0)}\right)d_{L-}^{(1)},
\end{align}
where $\widetilde{\delta\mu}\equiv \delta\mu \cdot {\rm sign}(d_{L-}^{(0)}-d_{L+}^{(0)})$, $r^{+}\equiv g_{L}^{-1}-d_{L+}^{(0)}$ and $r^{-}\equiv g_{L}^{-1}-d_{L-}^{(0)}$.
The derivation of eq. (\ref{expansion_dn_3}) is given in Appendix C.
\subsubsection{Phase transition}
\begin{figure}[h]
\begin{center}
\includegraphics[width=9cm]{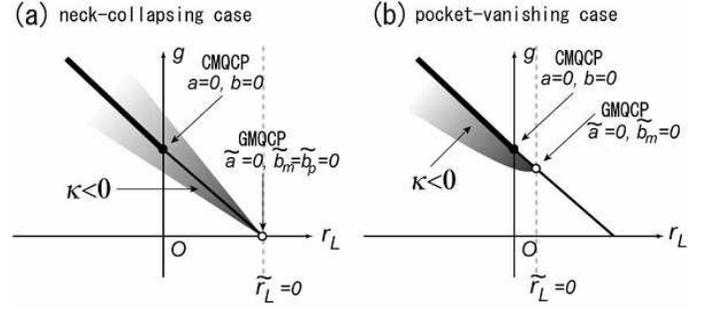}
\end{center}
\caption[]{(a)Schematic phase diagram for neck-collapsing case.
(b)Schematic phase diagram for pocket-vanishing case. In addition to MQCP of the canonical ensemble (CMQCP), MQCP of the grand canonical ensemble (GMQCP), defined by $\widetilde{a}=0$ and $\widetilde{b_{m}}=0$, appears. Within the shaded areas, the charge compressibility $\kappa$ becomes negative and the instability towards the electronic phase separation exists. For the pocket-vanishing case, the region where $\kappa$ becomes negative, only exists at the one side of the Lifshitz transition boundary, defined by $a=0$.
}
\label{GMQCP}
\end{figure}

For the neck-collapsing case, 
the charge compressibility $\kappa$ is scaled as
\begin{align}
\kappa = \frac{2}{g}\frac{r_{L}}{r_{L}-1},\label{kappa_rL_rL}
\end{align}
along the Lifshitz critical line. The derivation of eq. (\ref{kappa_rL_rL}) is given in Appendix C (see eq. (\ref{C15})).
Here, the inequality $r_{L}\le 1$ is always satisfied, leading to $\kappa <0$. The negative value of the charge compressibility $\kappa$ around the continuous transitions for $r_{L}>0$, means the instability toward the electronic phase separation. Then, the charge compressibility $\kappa$ is negative along the Lifshitz critical line, except for the case $g_{L}=0$. When $g_{L}$ approaches zero, from eq. (\ref{Lifhitz_canonical}), we see that $r_{L}\longrightarrow 1$, resulting in $\kappa \longrightarrow \infty$ from eq. (\ref{kappa_rL_rL}). In fact, the phase separation terminates just at $g_{L}=0$, which means that MQCP is located at $g_{L}=0$. At $g_{L} = 0$, the charge compressibility behaves as
\begin{align}
\kappa =\rho_{{\rm log}}\ln \frac{1}{|\delta n|},
\end{align}
as is derived in Appendix C (see eq. (\ref{C16})).
The free-energy expansion of $\widetilde{\mathcal{E}}$ around MQCP, namely around $g=0$ is given as 
\begin{align}
\widetilde{\mathcal{E}}\simeq& {\rm const.}-\delta\mu \delta n +\frac{1}{2\rho_{{\rm log}}}\frac{\delta n^{2}}{\ln \frac{1}{|\delta n|}},\notag\\
\simeq&{\rm const.}-\delta\mu \rho_{{\rm log}}\zeta\ln\frac{1}{|\zeta|} + \frac{\rho_{{\rm log}}}{2}\zeta^{2}\ln \frac{1}{|\zeta|},
\end{align}
which exhibits the same singularity mentioned in Table \ref{scLT}. Here we use the relation between $\delta n$ and $\zeta$ at $g_{L}=0$, namely $\delta n\simeq \rho_{{\rm log}}\zeta \ln 1/|\zeta|$, which is obtained from eq. (\ref{zeta_zeta}). Eventually, the marginal quantum criticality turns out to be nothing but the criticality of the noninteracting case. 
The endpoint of the phase separation only exists at $g_{L}=0$, as is illustrated in Fig. \ref{GMQCP} (a).
\par

\par
For the pocket-vanishing case, for $\delta\Delta=0$, $\kappa$ is given as
\begin{align}
\kappa =&(d_{L+}^{(0)}+d_{L-}^{(0)})\frac{\frac{1}{g_{L}}-\frac{2d_{L+}^{(0)}d_{L-}^{(0)}}{d_{L+}^{(0)}+d_{L-}^{(0)}}}{\frac{1}{g_{L}^{2}}-d_{L+}^{(0)}d_{L-}^{(0)}}\notag\\
=&
g_{L}\frac{d_{L+}^{(0)}+d_{L-}^{(0)}}{1+g_{L}\ \sqrt{d_{L+}^{(0)}d_{L-}^{(0)}}}\frac{r_{L}}{1-g_{L}\ \sqrt{d_{L+}^{(0)}d_{L-}^{(0)}}}.\label{kappa_d0_d1}
\end{align}
For the derivation of eq. (\ref{kappa_d0_d1}), see Appendix C (see eq. (\ref{last_kappa})).
Here the inequality $r_{L}\geq 1-g_{L}\ \sqrt{d_{L-}^{(0)}d_{L+}^{(0)}}$ always holds because the geometric average $\sqrt{d_{L-}^{(0)}d_{L+}^{(0)}}$ is greater than the harmonic average $2d_{L-}^{(0)}d_{L+}^{(0)}/(d_{L-}^{(0)}+d_{L+}^{(0)})$. Then the electronic phase separation inevitably occurs around MQCP of the canonical ensemble, where $r_{L}=0$. The endpoint of the electronic phase separation, which is defined by $1-g_{L}\ \sqrt{d_{L-}^{(0)}d_{L+}^{(0)}}=0$, also exists on the continuous Lifshitz transition line. 
The endpoint of the electronic phase separation, which is located at $\widetilde{a}=0$ and $\widetilde{b_{m}}=0$, is MQCP in the grand canonical ensemble.
The Lifshitz factor for the grand canonical ensemble, $\widetilde{r_{L}}$, is given as
\begin{align}
\widetilde{r_{L}}=1-g_{L}\ \sqrt{d_{L-}^{(0)}d_{L+}^{(0)}},
\end{align}
which is proportional to $\widetilde{b_{m}}$.
The instability towards the electronic phase separation signaled by the negative charge compressibility exists around MQCP of the canonical ensemble. We should note that the instability towards the phase separation only exists at the one side of the Lifshitz transition boundary of the canonical ensemble, where the pockets of the Fermi surface exist, as is illustrated in Fig. \ref{GMQCP} (b).
\subsubsection{Marginal quantum criticality}
The critical exponents $\beta$, $\delta$ and $\gamma$ for MQCP of the grand canonical ensemble are obtained as follows.
\paragraph{critical exponent; $\beta$}
The jump of the electron density $n$ at the first-order transition is given as the difference between the values of $\delta n$ at the two minima of the free-energy density, namely, $\delta n_{+}>0$ and $\delta n_{-}<0$. The exponent $\beta$ is obtained from the scaling $\delta n_{+}-\delta n_{-} \propto |\widetilde{b_{m}}|^{\beta}\propto |\widetilde{r_{L}}|^{\beta}$.
For the pocket-vanishing case, we obtain from eq. (\ref{expansion_dn_3}) that
\begin{align}
\widetilde{a}=
-\frac{|\widetilde{b_{m}}|^{2}}{4\widetilde{c_{m}}}+\mathcal{O}(\widetilde{b_{m}}^{3})
\end{align}
and $\widetilde{b_{m}}<0$, namely $\widetilde{r_{L}}<0$ determine the first-order transition boundary.
The jump of the electron density is given as
\begin{align}
\delta n_{+}-\delta n_{-} \simeq \frac{|\widetilde{b_{m}}|}{8\widetilde{c_{m}}}+\frac{|\widetilde{b_{m}}|^{2}}{8\widetilde{c_{m}}|\widetilde{b_{p}}|},
\end{align}
which indicates $\beta =1$.
\paragraph{critical exponent; $\delta$}
At the endpoint of the first-order transition boundary defined by $\widetilde{b_{m}}=0$,
$\widetilde{\delta \mu}$-dependence of $\delta n$ determines the critical exponent $\delta$ from the scaling $\delta n \propto \left|\widetilde{\delta \mu}\right|^{1/\delta}$. This relation is obtained from $\partial \widetilde{\mathcal{E}_{m}}/\partial \widetilde{\delta n} =0$ and $\partial \widetilde{\mathcal{E}_{p}}/\partial \widetilde{\delta n} =0$ at $\widetilde{b_{m}}=\widetilde{r_{L}}=0$ in eq. (\ref{expansion_dn_3}).
For the pocket-vanishing case, the behavior of $\delta n$ is given as
\begin{align}
|\delta n| = 
\left\{
\begin{array}{lc}
\sqrt{ \frac{|\widetilde{\delta \mu}|}{3c_{m}} }&(\widetilde{\delta \mu} < 0)\\
\frac{1}{2b_{p}}\widetilde{\delta \mu}&(\widetilde{\delta \mu} > 0)\label{eq_cross}
\end{array}
\right.,
\end{align}
which shows $\delta =2$ for $\widetilde{\delta \mu} < 0$ and $\delta =1$ for $\widetilde{\delta \mu} > 0$.
\paragraph{critical exponent; $\gamma$}
Upon approaching the endpoint $\widetilde{b_{m}}=\widetilde{r_{L}}=0$, the uniform charge compressibility $\kappa \equiv \left(\partial^{2}\widetilde{\mathcal{E}}/\partial \delta n^{2}\right)^{-1}$ is given from eq. (\ref{expansion_dn_3}) as
\begin{align}
\kappa=\left(2\widetilde{b_{\eta}}\right)^{-1}.
\end{align}
For $\widetilde{\delta n}<0$, the relation $\kappa \propto \left|\widetilde{b_{m}}\right|^{-\gamma}$ with $\gamma =1$ is satisfied. On the other hand, for $\widetilde{\delta n}>0$, $\kappa$ becomes constant value $d_{L-}^{(0)}$ resulting in $\gamma =0$.
In addition, as a function of $\delta n$ the charge compressibility $\kappa$ diverges as 
\begin{align}
\kappa \simeq \frac{1}{6c_{m}|\delta n|},
\label{scaling_kappa}
\end{align}
strictly at $\widetilde{b_{m}}=0$ for $\widetilde{\delta n} \rightarrow 0-$. The scaling (\ref{scaling_kappa}) is equivalent to $\delta n \propto \mu^{1/\delta}$ with the chemical potential $\mu\equiv \partial \widetilde{\mathcal{E}}/\partial \delta n$ and the exponent $\delta =2$.
In fact the exponent $\delta$ derived in eq. (\ref{eq_cross}) should be the same as $\delta$ derived here because $\widetilde{a}=-\widetilde{\delta \mu}$ is nothing but
the chemical potential $\mu$.
\par
These critical exponents for the pocket-vanishing case of the grand canonical ensemble are the same as those of MQCP of the canonical ensemble, which are shown in Table \ref{critical_exponent}.
\section{Numerical Studies}
In the present section, 
we demonstrate numerically the marginal quantum criticalities of the canonical ensemble. Here,
we consider a simple model for ferromagnets, namely $t$-$t'$ Hubbard model on square lattice defined by;
\begin{align}
\hat{H}=&\sum_{k\sigma}\varepsilon_{0}(k) c^{\dagger}_{k\sigma}c^{\ }_{k\sigma}+U\sum_{i}n_{i,\uparrow}n_{i,\downarrow},\label{hamiltonian}\\
\varepsilon_{0}(k)=&-2t(\cos k_{x}+\cos k_{y})+4t'\cos k_{x}
\cos k_{y},
\end{align}
where $c^{\dagger}_{k\sigma}$($c^{\ }_{k\sigma}$) represents the creation(annihilation) operator of an electron at wave number $k$ with spin $\sigma$. The nearest-neighbor and next-nearest-neighbor transfers are denoted by $t$ and $t'$, respectively. The onsite interaction is given by $U$ and the number operator of the $i$-th site is $n_{i\sigma}=c^{\dagger}_{i\sigma}c^{\ }_{i\sigma}$.
 It has been suggested, by a projection quantum Monte Carlo study, that, for $2t'/t > 0.55$, the $t$-$t'$ Hubbard model exhibits weak coupling ferromagnetic order, instead of anti-ferromagnetic one at the filling corresponding to the 
 van Hove singularity~\cite{Hlubina97}.
 In addition to the $t$-$t'$ Hubbard Hamiltonian (\ref{hamiltonian}), we introduce a transverse magnetic field $H_{\bot}$ as follows;
\begin{eqnarray}
H_{\bot}\sum_{i}\left(c^{\dagger}_{i\uparrow}c^{\ }_{i\downarrow}+c^{\dagger}_{i\downarrow}c^{\ }_{i\uparrow}\right),
\end{eqnarray}
where the Lande's $g$-factor and the Bohr magneton $\mu_{B}$ are included in the notation of $H_{\bot}$. Under a nonzero magnetic field $H_{\bot}$, the magnetization
\begin{eqnarray}
m_{\bot}\equiv N_{s}^{-1}\sum_{k}\langle c^{\dagger}_{k\downarrow}c^{\ }_{k\uparrow}\rangle
\end{eqnarray}
always remains finite.\par
To study the metamagnetic transitions within the present model, we use the mean-field approximation, and the terms of the on-site Coulomb repulsions are decoupled with the mean field $m_{\bot}$ as follows;
\begin{eqnarray}
&&-\frac{U}{N_{s}}\sum_{kq}\langle c^{\dagger}_{k\uparrow}c^{\ }_{k\downarrow}c^{\dagger}_{q\downarrow}c^{\ }_{q\uparrow} \rangle \nonumber\\
&\simeq & 
 -\sum_{k}\left(Um^{\ }_{\bot}c^{\dagger}_{k\uparrow}c^{\ }_{k\downarrow} + Um^{\ast}_{\bot} c^{\dagger}_{k\downarrow}c^{\ }_{k\uparrow}\right)
\nonumber\\
 &&+N_{s}U\left|m_{\bot}\right|^{2}.
\end{eqnarray}
Then, we obtain the quasiparticle dispersions of the spin-polarized states $\varepsilon_{\pm}(k)$ as,
\begin{eqnarray}
\varepsilon_{\pm}(k)=\varepsilon_{0}(k)\pm \left| Um_{\bot} + H_{\bot} \right|,
\end{eqnarray}
after the Bogoliubov transformation,
\begin{eqnarray}
 \alpha^{\dagger}_{k\pm}=\frac{1}{\sqrt{2}}c^{\dagger}_{k\uparrow} \mp \frac{1}{\sqrt{2}}c^{\dagger}_{k\downarrow}.
\end{eqnarray}
Here we assume that $m_{\bot}\cdot H_{\bot}>0$.
The electron density $n$ and $m_{\bot}$ are calculated by the following conditions;
\begin{align}
n=&\frac{1}{N_{s}}\sum_{k}\left\{
f(\varepsilon_{-}(k))+f(\varepsilon_{+}(k))
	\right\},\\
m_{\bot}
=&\frac{1}{2N_{s}}\sum_{k}\left\{
f(\varepsilon_{-}(k))-f(\varepsilon_{+}(k))
	\right\},
\end{align}
where $f(x)=\left(1+\exp \left[(x-\mu)/T\right]\right)^{-1}$ and the chemical potential $\mu$ and the magnetization $m_{\bot}$ must be determined self-consistently for the given electron density $n$.\par
\begin{figure}[h]
\begin{center}
\includegraphics[width=8cm]{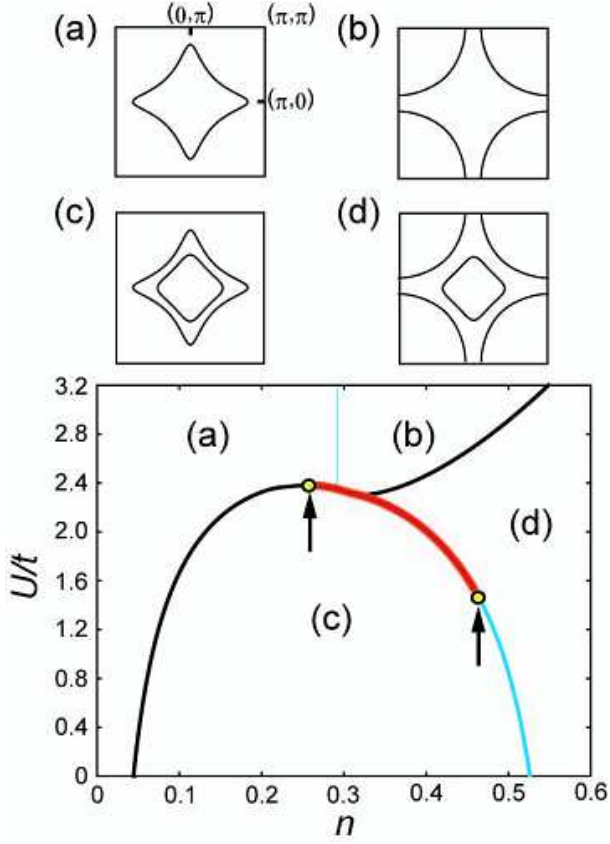}
\end{center}
\caption[]{Upper panel, (a),(b),(c) and (d): Shapes of Fermi surface corresponding to the regions in the phase diagram in the lower panel. Lower panel: Phase diagram of Lifshitz transitions at $t'/t=0.4$ and $H_{\bot}=0.06t$.
The thin black curves represent the continuous transition lines of the pocket-vanishing case. The thin blue curve represents the continuous transition line of the neck-collapsing case.
The thick red curve represents the first-order transition line. The vertical arrows indicate MQCPs. The phase boundary between (a) and (c) as well as between (b) and (d) represents the pocket-vanishing transition.
On the other hand the boundary between (c) and (d) as well as (a) and (b) represents the neck-collapsing transition.}
\label{Phase_diagram_LT}
\end{figure}

In analyzing the $t$-$t'$ Hubbard model with the external magnetic field by the mean-field approximation, 
we take $t'=0.4t$ as a typical value and leave $n$, $U$ and $H_{\bot}$ as control parameters. To avoid the instabilities toward commensurate antiferromagnetic orders, at least within the framework of the mean-field theory, we consider the low-electron-density region $n<0.6$.
The overall phase diagram is shown in Fig. \ref{Phase_diagram_LT}. For values of the electron density $n>0.29$, the Fermi level is located 
below the van Hove singularity of the majority spin sub-band $\varepsilon_{-}(k)$ at $U=0$. As $U$ increases, the Fermi level approaches the van Hove singularity and a neck-collapsing Lifshitz transition occurs. Above the neck-collapsing Lifshitz critical value of $U$, we further observe the pocket-vanishing Lifshitz transition.\par
On the other hand, for the lower electron density $n<0.29$, the Fermi level is always located below the van Hove singularity of the majority spin sub-band $\varepsilon_{-}(k)$ and only pocket-vanishing Lifshitz transitions can occur. These pocket-vanishing Lifshitz transitions at lower electron-densities are nothing but the transitions between partial and complete ferromagnetic states. For both of the neck-collapsing and the pocket-vanishing cases, the Lifshitz transitions contain both discontinuous and continuous parts. MQCPs appear at the boundaries between the first-order and continuous transition lines, as are shown in Fig. \ref{Phase_diagram_LT} as open circles.
\par
\begin{figure}[h]
\begin{center}
\includegraphics[width=7cm]{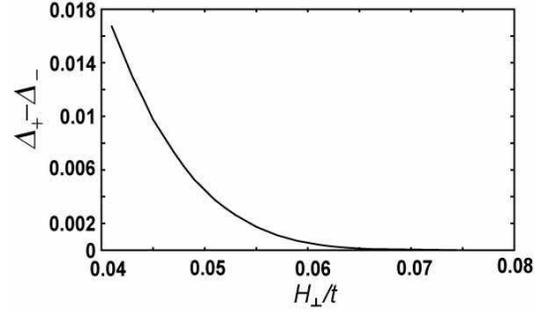}
\end{center}
\caption[]{Jump of mean field, $\delta\Delta_{+}-\delta\Delta_{-}$, as a function of $H_{\bot}/t$ at fixed electron-density $n=0.4$. As $H_{\bot}/t$ increases, the jump rapidly decreases to zero as a concave function of $H_{\bot}/t$.}
\label{beta_nc}
\end{figure}
\begin{figure}[h]
\begin{center}
\includegraphics[width=6cm]{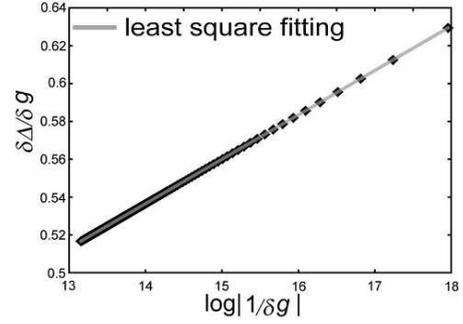}
\end{center}
\caption[]{Ratio of $\delta\Delta$ to $\delta g$ plotted as a function of $\ln 1/|\delta g|$ near MQCP at $U\simeq 1.478t, n\simeq 0.463$ for $H_{\bot}=0.06t$. The linear behavior of $\delta\Delta$ indicates that $\delta\Delta \propto \delta g\ln 1/|\delta g|$ is satisfied.}
\label{delta_nc}
\end{figure}

Let us consider at the electron density $n=0.4$, for example. Then the neck-collapsing Lifshitz transitions occur at the boundary between (c) and (d) in Fig. \ref{Phase_diagram_LT}. It becomes discontinuous for relatively small values of $H_{\bot}$. The jump of the mean field at the first-order transition rapidly decreases to zero with increasing $H_{\bot}$, as is shown in Fig. \ref{beta_nc}. The concave behavior of the jump $\delta\Delta_{+}-\delta\Delta_{-}$ as a function of $H_{\bot}/t$ indicates that the critical exponent $\beta$ should be larger than 1.
As shown in Fig. \ref{delta_nc}, near MQCP, $\delta\Delta$ behaves as $\delta\Delta \propto \delta g\ln 1/|\delta g|$. This scaling is consistent with eqs. (\ref{esse_sing}) and (\ref{delta_log}) indicating $\beta=\infty$ and $\delta =1+0$.

\begin{figure}[h]
\begin{center}
\includegraphics[width=6cm]{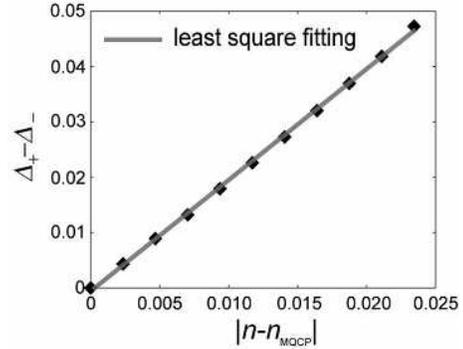}
\end{center}
\caption[]{Jump of mean field, $\delta\Delta_{+}-\delta\Delta_{-}$, as a function of $|n-n_{{\rm MQCP}}|$ at $H_{\bot}=0.06t$, where the critical electron density at MQCP is denoted by $n_{{\rm MQCP}}$. }
\label{beta_pv}
\end{figure}

\begin{figure}[h]
\begin{center}
\includegraphics[width=6cm]{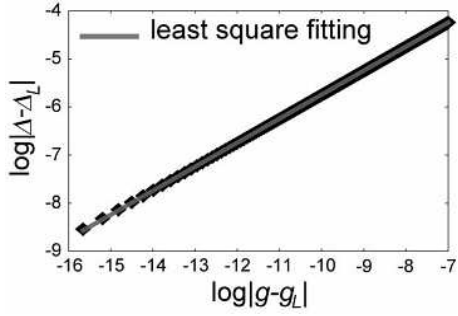}
\end{center}
\caption[]{Log-log plot of $|\delta\Delta|$ vs $|\delta g|$ around MQCP of pocket-vanishing case for negative $\delta g$ and $H_{\bot}=0.06t$. The relation $|\delta\Delta| \propto |\delta g|^{1/2}$ is obtained.}
\label{delta_pv}
\end{figure}

On the other hand, the density dependence of $\delta\Delta_{+}-\delta\Delta_{-}$ for the pocket-vanishing transition is shown in Fig. \ref{beta_pv}. Since the electron density measured from the critical electron density at MQCP $n_{{\rm MQCP}}$ is scaled linearly with $r_{L}$, it leads to $\beta =1$. As is shown in Fig. \ref{delta_pv}, the critical exponent for the pocket-vanishing case is given by $\delta =2$. These exponents are consistent with eqs. (\ref{beta_linear}) and (\ref{delta_sq}).

\section{Summary and Discussion}
In the present paper, electron correlation effects on Lifshitz transitions have been investigated 
by assuming the presence of a broken-symmetry order. Through the interplay with the order parameter amplitude, electron correlation causes the first-order transition in contrast with the noninteracting case. The first-order transition boundary appears in a part of the parameter space while a continuous transition boundary also exists. These two boundaries meet at the marginal quantum critical point (MQCP) as seen in Fig. \ref{marginal}. The continuous boundary at $T=0$ appears because this transition is a topological one. The topological nature deeply influences the quantum criticality and the criticality does not follow the Ginzburg-Landau-Wilson (GLW) scheme.
\par
Let us summarize the parameters to be controlled here. The Fermi level $\zeta$ measured from the critical point is the control parameter in the noninteracting case. In the present interacting system, the primary control parameter may alternatively be taken as $\delta\Delta$, the order parameter of the broken symmetry. Then $\zeta$ is determined from $\delta\Delta$ self-consistently. In terms of the experimental control, $\delta\Delta$ is self-consistently determined from an external control parameter such as the interaction parameter $\delta g$ measured from the critical point, which may be directly expressed by pressure, for example. Another controllable parameter is the Lifshitz factor $r_{L}$ as is seen in Fig. \ref{marginal}. The Lifshitz factor $r_{L}$ is considered to be controlled by the external magnetic field $H_{\bot}$ and/or the electron density $n$.\par
We first summarize the case of the canonical ensemble. In the presence of interactions, the singularities of the continuous Lifshitz transitions seen in
the susceptibilities $\chi$ are different from the noninteracting case shown in Table \ref{scLT}. For $r_{L}=0$, the susceptibilities $\chi$ behave as
\begin{eqnarray}
\chi\propto \left\{
\begin{array}{cl}
\ln \frac{1}{|\delta\Delta|}& ({\rm neck\verb|-| collapsing})\\
|\delta\Delta|^{-1}& ({\rm pocket\verb|-| vanishing})
\end{array}
\right. ,
\end{eqnarray}
which corresponds to eq. (\ref{chi_D}).
In terms of the control parameter $\delta g$, the susceptibility $\chi$ behaves as
\begin{eqnarray}
\chi\propto \left\{
\begin{array}{cl}
\ln \frac{1}{\delta g}& ({\rm neck\verb|-| collapsing})\\
\delta g^{-1/2}& ({\rm  pocket\verb|-| vanishing})
\end{array}
\right. .
\end{eqnarray}
In terms of $r_{L}$, the susceptibility $\chi$ behaves as
\begin{eqnarray}
\chi &\propto& r_{L}^{-1}.
\end{eqnarray}
In contrast to the noninteracting case, where the susceptibilities do not diverge, the marginal quantum criticality exhibits the diverging fluctuation, especially for the pocket-vanishing case. We also note that the shift of the Fermi level $\zeta$ behaves as
\begin{eqnarray}
\zeta \propto \left\{
\begin{array}{@{\,}c@{\ }l@{}}
-\delta\Delta& ({\rm neck\verb|-| collapsing})\\
 -\frac{d_{L-}^{(0)}-d_{L+}^{(0)}}{d_{L-}^{(0)}+d_{L+}^{(0)}}\delta\Delta&({\rm  pocket\verb|-| vanishing}, \delta\Delta <0)\\
-\delta\Delta& ({\rm  pocket\verb|-| vanishing}, \delta\Delta>0)
\end{array}
\right.,
\end{eqnarray}
when $\delta\Delta \longrightarrow 0$.
\par
Next we summarize the grand canonical case. For the neck-collapsing case, the electronic phase separation always occurs around the Lifshitz critical line, in the presence of the interaction. On the other hand, for the pocket-vanishing case, the endpoint of the electronic phase separation, which is defined by $g\ \sqrt{d_{0}\widetilde{d}_{0}}=1$, exists. Then the charge compressibility $\kappa$ diverges as
\begin{eqnarray}
\kappa \propto |\delta\Delta|^{-1}\propto |\delta n|^{-1}\propto |\delta \mu|^{-1/2}.
\end{eqnarray}
We also control the divergence of $\kappa$ around grand canonical MQCP by changing coupling constant $\delta g$, which turns out to be proportional to $\delta \mu$.
In terms of $r_{L}$, $\kappa$ is scaled as
\begin{eqnarray}
\kappa \propto \frac{r_{L}}{\widetilde{r_{L}}},
\end{eqnarray}
where $\widetilde{r_{L}}=1-g\ \sqrt{d_{0}\widetilde{d}_{0}}$.
At the grand canonical MQCP where $\widetilde{r_{L}} \longrightarrow 0$, $\kappa$ diverges as $\widetilde{r_{L}}^{-1}$. In this case, the shift of the Fermi level $\zeta$ behaves as
\begin{eqnarray}
\zeta \propto \delta n,
\end{eqnarray}
when $\delta n \longrightarrow 0$.
\par
The macroscopic electronic phase separation is prevented by the charge neutrality, in the metallic compounds. However, the instability towards the phase separation itself exist intrinsically and the uniform state is unstable anyhow. This instability around MQCP of the grand canonical case competes with the charge neutrality. Then the competition would result in the electronic inhomogeneity~\cite{Lorenzana}. The existence of the electronic inhomogeneity has been discussed for the HTSCs~\cite{Pan} and the manganites~\cite{Moreo} and has attracted increasing interest. The electronic phase separation induced by the two-dimensional van Hove singularity, which is nothing but the phase separation induced by the neck-collapsing Lifshitz criticality, was also discussed in the context of the multiple metamagnetic transition observed in the bilayer Ruthenates~\cite{Honerkamp}.
\par
For both of the canonical and grand canonical ensembles, we obtain the critical exponents of MQCP, which are different from those of the mean-field Ising universality, as is summarized in Table \ref{critical_exponent}. It is necessary to further examine the validity of these exponents, since these critical exponents are obtained by the mean-field theory, which ignores spatial and temporal fluctuations around critical points. For the neck-collapsing case, because of the logarithmic singularity, it may be difficult to determine the critical exponents of MQCP in real systems although the first-order jump may be observed. On the other hand, for the pocket-vanishing case, the power-law singularities may easily be observed.\par
The free-energy expansion and the mean-field exponents for MQCP of the pocket-vanishing case may be justified beyond the mean-field approximation. This is because the dynamical exponent $z$ is expected to be equal to $4$. In two dimensions, the effective dimension of the marginal quantum critical phenomena of the pocket-vanishing case, $d+z$, may be at the upper critical dimension of the $\varphi^{3}$-theory, namely $d_{c}=6$~\cite{Misawa06}. The details will be reported in a separate publication.
\par
Although we consider the ferromagnetic band dispersion (\ref{dispersion}) and the corresponding self-consistent equation (\ref{SCE1}) for simplicity, the essential physics of the marginal quantum criticality is retained
if individual band dispersions are considered in other possible cases. The ferromagnetic model treated in this paper can in fact also be mapped to another model, namely $t$-$t'$ Hubbard model on two layers which contains homogeneously coupled two-dimensional square lattices. Then the spin degrees of freedom in the ferromagnetic model are mapped to the layer degrees of freedom. These two degrees of freedom are referred to as A and B subbands in \S 2. The external magnetic field is mapped to an inter-layer transfer integral. Additional spin degrees of freedom exist in the bilayer model. When the magnetic orders are absent, the susceptibility $\chi$ here represents the inter-layer fluctuation of the electron density. This bilayer model is expected to have common features to the layered two-dimensional metals like Bi-2212.
\section*{Acknowledments}
This work has been supported from the Grant in Aid for Scientific Research on Priority Area under the grant numbers 17071003 and 16076212 from the Ministry of Education, Culture, Sports, Science and Technology. A part of our computation in this work has been done using the facilities of the Supercomputer Center, Institute for Solid State Physics, University of Tokyo.
\setcounter{section}{1}
\setcounter{equation}{0}
\renewcommand\theequation{\Alph{section}.\arabic{equation}}
\section*{Appendix A: Derivation of free-energy expansion of canonical ensemble}
\paragraph{neck-collapsing case}
For the band structure 
$\varepsilon_{\pm}(k,\Delta)=\varepsilon_{0}(k)\pm \left(\Delta_{0}+\Delta\right)$, the density of states $D_{\pm}(\varepsilon,\Delta_{L}+\delta\Delta)$ can be replaced by
\begin{eqnarray}
D_{\pm}(\varepsilon-\varepsilon_{F}^{L}\mp\delta\Delta)\equiv
D_{\pm}(\varepsilon,\Delta_{L}+\delta\Delta),
\end{eqnarray}
where the right hand side is defined in eqs. (\ref{D_+d}) and (\ref{D_-d}). Here, we assume $\Delta_{0}\cdot\Delta >0$. Then the condition for the fixed electron density (\ref{eldd}) can be rewritten as
\begin{eqnarray}
n=\int_{-\infty}^{\zeta +\delta\Delta}d\zeta_{-}D_{-}(\zeta_{-})
+\int_{-\infty}^{\zeta -\delta\Delta}d\zeta_{+}D_{+}(\zeta_{+}),
\end{eqnarray}
where $\zeta_{\pm}$ is given as $\zeta_{\pm}=\varepsilon-\varepsilon_{F}^{L}\mp\delta\Delta$. The critical electron density at the Lifshitz transition $n_{L}$ is defined as
\begin{eqnarray}
n_{L}=\int_{-\infty}^{0}d\zeta_{-}D_{-}(\zeta_{-})
+\int_{-\infty}^{0}d\zeta_{+}D_{+}(\zeta_{+}).
\end{eqnarray}
Then the condition $n=n_{L}$ results in
\begin{align}
0=&n-n_{L}\notag\\
=&\int_{0}^{\zeta +\delta\Delta}d\zeta_{-}D_{-}(\zeta_{-})
+\int_{0}^{\zeta -\delta\Delta}d\zeta_{+}D_{+}(\zeta_{+}).\label{zero_deln}
\end{align}
For the neck-collapsing case, the expansions of the density of states (\ref{D_+d}) and (\ref{D_-d}) are rewritten as
\begin{align}
D_{+}(\zeta_{+})=& d_{L+}^{(0)}(\Delta_{L})+d_{L+}^{(1)}(\Delta_{L})\zeta_{+}+\mathcal{O}\left(\zeta_{+}^{2}\right),\label{D_+d_s}\\
D_{-}(\zeta_{-})=& d_{L-}^{(0)} + \rho_{{\rm log}}\ln \frac{1}{\left|\zeta_{-}\right|}+d_{L-}^{(0)}\zeta_{-}+\mathcal{O}\left(\zeta_{-}^{2}\right).\label{D_-d_s}
\end{align}
By substituting the expansions (\ref{D_+d_s}) and (\ref{D_-d_s}) into (\ref{zero_deln}), we obtain an approximate formula for the constraint of the fixed electron density as
\begin{align}
\rho_{{\rm log}}\ln \frac{1}{\left|\zeta +\delta\Delta \right|}\left(\zeta +\delta\Delta \right)+\left({d_{L-}^{(0)}}+\rho_{{\rm log}}\right)\left(\zeta +\delta\Delta\right)& \notag\\
+d_{L+}^{(0)}\left(\zeta-\delta\Delta\right)
+\mathcal{O}(\delta\Delta^{2}) =& 0,\label{conservation}
\end{align}
up to the order of $\delta\Delta^{2}$.
To obtain the solution for eq. (\ref{conservation}), we may take the following form of $\zeta$,
\begin{eqnarray}
\zeta=-\delta\Delta (1-\mathcal{R}),\label{Rdef}
\end{eqnarray}
where $\mathcal{R}$ is a function of $\delta\Delta$ and is defined by this relation. From eq. (\ref{conservation}), it is apparent that $\delta\Delta$ is proportional to $\zeta$ except for a logarithmic correction so that $\mathcal{R}$ may have an amplitude of order unity.
Now eq.(\ref{conservation}) can be rewritten by using (\ref{Rdef}) as
\begin{align}
\rho_{{\rm log}}\delta\Delta \mathcal{R} \ln \frac{1}{\delta\Delta \mathcal{R}} +(d_{L+}^{(0)}+\rho_{{\rm log}}) \delta\Delta \mathcal{R}&\notag\\ 
-d_{L+}^{(0)}\left(2\delta\Delta-\delta\Delta \mathcal{R}\right)+\mathcal{O}(\delta\Delta^{2})=&0.\label{conserve2}
\end{align}
We can drop linear terms of $\delta\Delta \mathcal{R}$ to discuss the asymptotic behavior, since $\delta\Delta \mathcal{R}$ is higher order than $\delta\Delta \mathcal{R} \ln \frac{1}{\delta\Delta \mathcal{R}}$. 
Then eq. (\ref{conserve2}) is simplified as follows;
\begin{eqnarray}
\mathcal{R}\simeq\frac{2d_{L+}^{(0)}}{\rho_{{\rm log}}}\frac{1}{\ln \frac{1}{\delta\Delta \mathcal{R}}}.\label{iterative}
\end{eqnarray}
The asymptotic solution of eq. (\ref{conservation}) is accordingly obtained by solving (\ref{iterative}) iteratively as
\begin{eqnarray}
\zeta \sim -\delta \Delta \left(1- \frac{2d_{L+}^{(0)}}{\rho_{{\rm log}}}\frac{1}{\ln \frac{1}{\left|\delta\Delta\right|}}\right),\label{delta_mu_s}
\end{eqnarray}
which completes the derivation of eq. (\ref{delta_mu}). The presence of the logarithmic divergence due to the van Hove singularities makes the asymptotic behavior determined only by $d_{L+}^{(0)}$ and $\rho_{{\rm log}}$.
The free-energy density at $T=0$ given by eq. (\ref{fedr}) is rewritten as
\begin{align}
\mathcal{E}_{0}=&\int_{-\infty}^{\zeta+\delta\Delta}d\zeta_{-}\left(\zeta_{-}-\delta\Delta\right)D_{-}(\zeta_{-})\notag\\
&+\int_{-\infty}^{\zeta-\delta\Delta}d\zeta_{+}\left(\zeta_{+}+\delta\Delta\right)D_{+}(\zeta_{+})+\frac{\left(\Delta_{L}+\delta\Delta\right)^{2}}{g},\notag\\
=&E_{-L}+E_{+L}-\left(n_{-L}-n_{+L}\right)\delta\Delta+\frac{\left(\Delta_{L}+\delta\Delta\right)^{2}}{g}\notag\\
&+\int_{0}^{\zeta+\delta\Delta}d\zeta_{-}\left(\zeta_{-}-\delta\Delta\right)D_{-}(\zeta_{-})\notag\\
&+\int_{0}^{\zeta-\delta\Delta}d\zeta_{+}\left(\zeta_{+}+\delta\Delta\right)D_{+}(\zeta_{+}),
\label{fred_zero}
\end{align}
where the constant terms $E_{\pm L}$ and $n_{\pm L}$ are defined as $E_{\pm L}\equiv \int_{-\infty}^{0}d\zeta_{\pm}\zeta_{\pm}D_{\pm}(\zeta_{\pm})$ and $n_{\pm L}\equiv \int_{-\infty}^{0}d\zeta_{\pm}D_{\pm}(\zeta_{\pm})$.\par
By substituting the expansions (\ref{D_+d_s}) and (\ref{D_-d_s}) and the relation (\ref{delta_mu_s}) into (\ref{fred_zero}), the expansion of $\mathcal{E}_{0}$ around the Lifshitz critical point is obtained as eq. (\ref{free_energy_nc}).
%
\paragraph{pocket-vanishing case}
For the pocket-vanishing case, the expansions of the density of states (\ref{D_+d_pvs}) and (\ref{D_-d_pvs}) are rewritten as
\begin{align}
D_{+}(\zeta_{+})=&\theta (\zeta_{+})\left[{d_{L+}^{(0)}}+{d_{L+}^{(1)}}\left(\zeta_{+}\right)+\mathcal{O}\left(\zeta_{+}^{3}\right)\right],\label{D_+d_pvss}\\
D_{-}(\zeta_{-})=&d_{L-}^{(0)}+d_{L-}^{(1)}\left(\zeta_{-}\right)
+\mathcal{O}\left(\zeta_{-}\right)^{3}
.\label{D_-d_pvss}
\end{align}
By substituting the expansions (\ref{D_+d_pvss}) and (\ref{D_-d_pvss}) into (\ref{zero_deln}), we obtain an approximate formula for the constraint of the fixed electron density as
\begin{align}
0=&(d_{L-}^{(0)}+d_{L+}^{(0)})\zeta + (d_{L-}^{(0)}-d_{L+}^{(0)})\delta\Delta\notag\\
&-\frac{1}{2}(d_{L-}^{(1)}-d_{L+}^{(1)})\left(\zeta^{2}+\delta\Delta^{2}\right)\notag\\
&-(d_{L-}^{(1)}+d_{L+}^{(1)})\zeta\delta\Delta+\mathcal{O}(\delta\Delta^{3})
,\label{constraint_pv}
\end{align}
where we assume $\zeta-\delta\Delta>0$, otherwise the constraint of the fixed electron density is given as $0=d_{L-}^{(0)}(\zeta+\delta\Delta)+\frac{1}{2}d_{L-}^{(1)}(\zeta+\delta\Delta)^{2}$. From eq. (\ref{constraint_pv}), the relation
\begin{align}
\zeta =&-\frac{d_{L-}^{(0)}-d_{L+}^{(0)}}{d_{L-}^{(0)}+d_{L+}^{(0)}}\delta\Delta+2
\frac{d_{L-}^{(1)}{d_{L+}^{(0)}}^{2}-d_{L+}^{(1)}{d_{L-}^{(0)}}^{2}}{\left(d_{L-}^{(0)}+d_{L+}^{(0)}\right)^{3}}\delta\Delta^{2}\notag\\
&+\mathcal{O}\left(\delta\Delta^{3}\right),\label{delmu_pv}
\end{align}
for $\zeta-\delta\Delta=-\frac{2d_{L-}^{(0)}}{d_{L-}^{(0)}+d_{L+}^{(0)}}\delta\Delta >0$, otherwise $\zeta=-\delta\Delta$ is obtained.\par
By substituting the expansions (\ref{D_+d_pvss}) and (\ref{D_-d_pvss}) and the relation (\ref{delmu_pv}) into (\ref{fred_zero}), the expansion of $\mathcal{E}_{0}$ around the Lifshitz critical point is obtained as eq. (\ref{free_energy_pv}).
\setcounter{section}{2}
\setcounter{equation}{0}
\section*{Appendix B: Critical exponents of neck-collapsing case}Because of the van Hove singularity, the critical exponents are not well-defined on the marginal quantum critical point of the neck-collapsing case. The relation, $\delta g \propto \delta\Delta/\ln\frac{1}{\delta\Delta}$, has to correspond to $\delta g \propto \delta\Delta^{\delta}$. Here, to focus on the slow convergence of $1/\ln\frac{1}{\delta\Delta}$ as  $\delta\Delta$ approaches zero, we use a symbolic notation as $1/\ln\frac{1}{\delta\Delta}\propto \delta\Delta^{+0}$. This notation represents that $1/\ln\frac{1}{\delta\Delta}$ converges to zero more slowly than $\delta\Delta^{p}$, where the exponent $p$
takes the limit $+0$.
From eq.(\ref{esse_sing}), $\delta\Delta$ depends on $r_{L}<0$ as the following essentially singular form, 
\begin{eqnarray}
|\delta\Delta| \propto \exp \left[ - {\rm const.} \times |r_{L}|^{-1}\right],
\label{4_delta}
\end{eqnarray}
when $r_{L}$ approaches zero along the first-order transition boundary.
Equation (\ref{4_delta}) yields the following relation, 
\begin{eqnarray}
|r_{L}|\propto \frac{1}{\ln 1/\delta\Delta}.
\end{eqnarray}
Then, by using the above notation, the relation $|r_{L}|\propto |\delta\Delta|^{+0}$ holds symbolically. It may be rewritten as $|\delta\Delta|\propto |r_{L}|^{1/+0}$ symbolically.
On the other hand, because the relation $\left(\ln 1/\delta\Delta\right)^{-1}\propto \delta g/\delta\Delta$ or the symbolic form $|\delta\Delta|^{+0}\propto \delta g/\delta\Delta$ holds from eq. (\ref{delta_log}), another relation is obtained as
\begin{eqnarray}
|r_{L}|\propto \frac{\delta g}{\delta \Delta},
\end{eqnarray}
which corresponds to a contract $+0/+0=1$ for the symbolic notation, because of the symbolic relation $|\delta\Delta|^{+0}\propto\left||r_{L}|^{1/+0}\right|^{+0} \propto \delta g/\delta\Delta$.
Then critical exponents are expressed symbolically as 
\begin{eqnarray}
\beta=\frac{1}{+0},\ \delta = 1+0.
\end{eqnarray}
By noting the fact that the critical exponent $\gamma$ is equal to 1, the scaling relation, $\beta ( \delta -1 )= \gamma$ holds with the contract $+0/+0=1$.
\setcounter{equation}{0}
\setcounter{section}{3}
\section*{Appendix C: Derivation of free-energy expansion of grand canonical ensemble}To calculate
the free-energy expansion for the grand canonical ensemble, we assume the difference between $n$ and $n_{L}$ as $\delta n\equiv n-n_{L}$. From eq. (\ref{zero_deln}), $\delta n$ is given as
\begin{eqnarray}
\delta n=\int_{0}^{\zeta +\delta\Delta}d\zeta_{-}D_{-}(\zeta_{-})
+\int_{0}^{\zeta -\delta\Delta}d\zeta_{+}D_{+}(\zeta_{+}),\label{deldel_n}
\end{eqnarray}
where $\delta\Delta$ and $\zeta$ are functions of $\delta n$. For the fixed coupling constant $g=g_{L}$, the mean field $\delta\Delta$ satisfies the following self-consistent equation
\begin{align}
\Delta_{L}+\delta\Delta=\frac{g_{L}}{2}\left\{
	\int_{-\infty}^{\zeta+\delta\Delta}d\zeta_{-}D_{-}(\zeta_{-})-
	\int_{-\infty}^{\zeta-\delta\Delta}d\zeta_{+}D_{+}(\zeta_{+})
	\right\},
\end{align}
where $\Delta_{L}$ is given as
\begin{eqnarray}
\Delta_{L}=\frac{g_{L}}{2}\left\{
	\int_{-\infty}^{0}d\zeta_{-}D_{-}(\zeta_{-})-
	\int_{-\infty}^{0}d\zeta_{+}D_{+}(\zeta_{+})
	\right\}.
\end{eqnarray}
Then $\delta\Delta$ is given as
\begin{eqnarray}
\delta\Delta =\frac{g_{L}}{2}\left\{
	\int_{0}^{\zeta+\delta\Delta}d\zeta_{-}D_{-}(\zeta_{-})-
	\int_{0}^{\zeta-\delta\Delta}d\zeta_{+}D_{+}(\zeta_{+})
	\right\}.\label{deldel_d}
\end{eqnarray}
\paragraph{neck-collapsing case} 
The free-energy density is given as
\begin{align}
\mathcal{E}=&\int_{0}^{\zeta+\delta\Delta}d\zeta_{-}
\left(\zeta_{-}-\delta\Delta\right)D_{-}(\zeta_{-})\notag\\
&+\int_{0}^{\zeta-\delta\Delta}d\zeta_{+}
\left(\zeta_{+}+\delta\Delta\right)D_{+}(\zeta_{+})\notag\\
&+\frac{\delta\Delta}{g}+\frac{gn^{2}}{4},\notag\\
=&\frac{1}{2}\left(d_{L-}^{(0)}+{d_{L+}^{(0)}}\right)\left(\zeta^{2}-\delta\Delta^{2}\right)\notag\\
&+\frac{\rho_{{\rm log}}}{2}(\zeta^{2}-\delta\Delta^{2})\ln\frac{1}{|\zeta + \delta\Delta|}+\frac{\rho_{{\rm log}}}{4}(\zeta +\delta\Delta)(\zeta -3\delta\Delta)\notag\\
&+\frac{\delta\Delta}{g}+\frac{gn^{2}}{4}+\mathcal{O}(\delta\Delta^{3}).\label{ast_free_energy}
\end{align}
By substituting the expansions (\ref{D_+d_s}) and (\ref{D_-d_s}) into (\ref{deldel_n}) and (\ref{deldel_d}), we obtain the following relations
\begin{align}
\delta n=&({d_{L-}^{(0)}}+\rho_{{\rm log}})\left(\zeta +\delta\Delta\right)
+\rho_{{\rm log}}\left(\zeta +\delta\Delta\right)\ln\frac{1}{\left|\zeta +\delta\Delta\right|}\notag\\
&+d_{L+}^{(0)}\left(\zeta-\delta\Delta\right)+\mathcal{O}(\delta\Delta^{2}),\label{deldel_n_s}\\
\frac{2}{g_{L}}\delta\Delta=&({d_{L-}^{(0)}}+\rho_{{\rm log}})\left(\zeta +\delta\Delta\right)
+\rho_{{\rm log}}\left(\zeta +\delta\Delta\right)\ln\frac{1}{\left|\zeta +\delta\Delta\right|}\notag\\
&-d_{L+}^{(0)}\left(\zeta-\delta\Delta\right)+\mathcal{O}(\delta\Delta^{2}).\label{deldel_d_s}
\end{align}
From eqs. (\ref{deldel_n_s}) and (\ref{deldel_d_s}), we obtain the relation which does not include logarithmic terms as
\begin{eqnarray}
\delta n -\frac{2}{g}\delta\Delta=2d_{L+}^{(0)}\left(\zeta-\delta\Delta\right)+\mathcal{O}(\delta\Delta^{2}).\label{n_del_mu}
\end{eqnarray}
By solving (\ref{n_del_mu}), $\delta\Delta$ is given as
\begin{eqnarray}
\delta\Delta=\frac{\delta n-2d_{L+}^{(0)}\zeta}{\frac{2}{g_{L}}-2d_{L+}^{(0)}}+\mathcal{O}(\delta n^{2}).\label{C_8}
\end{eqnarray}
Then we can eliminate $\delta\Delta$ from (\ref{deldel_n_s}) as
\begin{eqnarray}
\delta n&=&({d_{L-}^{(0)}}+\rho_{{\rm log}})\frac{g_{L}\delta n+2r_{L}\zeta}{2s}\nonumber\\
&&+\rho_{{\rm log}}\frac{g\delta n+2r_{L}\zeta}{2s}\ln \frac{2s}{\left|g_{L}\delta n+2r_{L}\zeta\right|}
\nonumber\\&&
+d_{L+}^{(0)}\frac{-g_{L}\delta n+2\zeta}{2s}+\mathcal{O}(\delta n^{2}),\label{dn_dmu_nc_long}
\end{eqnarray}
where $r_{L}=1-2g_{L}d_{L+}^{(0)}$ and $s=1-g_{L}d_{L+}^{(0)}$. To obtain the solution for eq.(\ref{dn_dmu_nc_long}), we may take the following form of $\delta n$,
\begin{eqnarray}
\delta n =-\frac{2r_{L}\zeta}{g_{L}}\left(1+\mathcal{S}\right),\label{mathcal_S}
\end{eqnarray}
where $\mathcal{S}$ is a function of $\zeta$ and is defined by this relation. The relation (\ref{dn_dmu_nc_long}) can be rewritten by using (\ref{mathcal_S}) as
\begin{eqnarray}
-\frac{2r_{L}\zeta}{g}\simeq
2d_{L+}^{(0)}\zeta-\rho_{{\rm log}}\frac{r_{L}\zeta\mathcal{S}}{s}\ln\frac{s}{\left|r_{L}\zeta\mathcal{S}\right|},
\end{eqnarray}
where the terms, which are order of $\zeta \mathcal{S}$, are dropped. Then the asymptotic form of $\mathcal{S}$ is given as
\begin{eqnarray}
\mathcal{S}\sim \frac{2}{g_{L}\rho_{{\rm log}}}\frac{s^{2}}{r_{L}}\frac{1}{\ln\frac{1}{|\zeta|}}.
\end{eqnarray}
Then, eq. (\ref{mathcal_S}) can be rewritten as
\begin{eqnarray}
\delta n \simeq -\frac{2r_{L}\zeta}{g_{L}}\left\{
1+\frac{2}{g_{L}\rho_{{\rm log}}}\frac{s^{2}}{r_{L}}
\frac{1}{\ln\frac{1}{|\zeta|}}	\right\}.\label{deln_delmu_ncnc}
\end{eqnarray}
We can also solve eq. (\ref{deln_delmu_ncnc}) iteratively as
\begin{eqnarray}
\zeta \simeq -\frac{g_{L}\delta n}{2r_{L}}
\left(1-\frac{2s^{2}}{g_{L}\rho_{{\rm log}}r_{L}}
\frac{1}{\ln \frac{1}{|\delta n|}}\right),\label{deln_delmu_ncnc_1}
\end{eqnarray}
where we assume $r_{L}\neq 0$. 
%
%
By substituting (\ref{deln_delmu_ncnc_1}) into (\ref{C_8}), we obtain $\delta\Delta$ as a function of $\delta n$ as
\begin{align}
\delta\Delta \simeq \frac{g_{L}}{2r_{L}}\delta n -\frac{g_{L}d_{L+}^{(0)}s}{\rho_{{\rm log}}r_{L}^{2}}\frac{\delta n}{\ln \frac{1}{|\delta n|}}.\label{sharp}
\end{align}
By substituting (\ref{deln_delmu_ncnc_1}) and (\ref{sharp}) into (\ref{ast_free_energy}), we obtain the free-energy expansion with respect to $\delta n$ as
\begin{eqnarray}
\mathcal{E}_{\eta}-\frac{g_{L}n_{L}}{2}\delta n \simeq {\rm const.}
+\widetilde{b_{\eta}}\delta n^{2}+\widetilde{c_{\eta}}\frac{\delta n^{2}}{\ln \frac{1}{|\delta n|}},\label{sharp_free_energy}
\end{eqnarray} 
where $\eta =p,m$ and the coefficients are defined as
\begin{align}
&\widetilde{b_{m}}=\widetilde{b_{p}}=-\frac{g_{L}}{4}\left(\frac{1-r_{L}}{r_{L}}\right),\\
&\widetilde{c_{m}}=\widetilde{c_{p}}=-\left(d_{L-}^{(0)}+d_{L+}^{(0)}+\rho_{{\rm log}}-g_{L}^{-1}\right)\frac{sg_{L}}{2\rho_{{\rm log}}r_{L}^{2}}.
\end{align}
From eq. (\ref{sharp_free_energy}), 
the uniform charge compressibility $\kappa$ is given as
\begin{align}
\kappa^{-1} \simeq& \frac{\partial^{2}}{\partial \delta n^{2}}( \mathcal{E}_{\eta}-\frac{g_{L}n_{L}}{2})\notag\\
\simeq&\frac{g}{2}-\frac{g}{2r_{L}}
\left(1-\frac{2s^{2}}{g\rho_{{\rm log}}r_{L}}
\frac{1}{\ln \frac{1}{|\delta n|}}\right).\label{C15}
\end{align}
In this case, from the inequality $r_{L}=1-g_{L}d_{L+}^{(0)}\le 1$, the charge compressibility $\kappa$ is negative along the Lifshitz critical line, except for the noninteracting case. For $g = 0$, the charge compressibility behaves as
\begin{eqnarray}
\kappa =\rho_{{\rm log}}\ln \frac{1}{|\delta n|}.\label{C16}
\end{eqnarray}
\paragraph{pocket-vanishing case}
For $\zeta-\delta\Delta >0$, the free-energy density is given as
\begin{align}
\mathcal{E}=&\int_{0}^{\zeta+\delta\Delta}d\zeta_{-}
\left(\zeta_{-}-\delta\Delta\right)D_{-}(\zeta_{-})\notag\\
&+\int_{0}^{\zeta-\delta\Delta}d\zeta_{+}
\left(\zeta_{+}+\delta\Delta\right)D_{+}(\zeta_{+})\notag\\
&+\frac{\delta\Delta}{g}+\frac{gn^{2}}{4},\notag\\
=&\frac{1}{2}\left(d_{L-}^{(0)}+{d_{L+}^{(0)}}\right)\left(\zeta^{2}-\delta\Delta^{2}\right)\notag\\
&+\frac{d_{L-}^{(1)}+{d_{L+}^{(1)}}}{3}\zeta^{3}+\frac{1}{2}\left(d_{L-}^{(1)}-{d_{L+}^{(1)}}\right)\zeta^{2}\delta\Delta\notag\\
&-\frac{1}{6}\left(d_{L-}^{(1)}-{d_{L+}^{(1)}}\right)\delta\Delta^{3}+\frac{\delta\Delta}{g}+\frac{gn^{2}}{4}+\mathcal{O}(\delta\Delta^{4}).\label{EEpv}
\end{align}
By substituting the expansions (\ref{D_+d_pvss}) and (\ref{D_-d_pvss}) into (\ref{deldel_n}) and (\ref{deldel_d}), we obtain the following relations
\begin{align}
\delta n=&d_{L-}^{(0)}(\zeta+\delta\Delta)+\frac{d_{L-}^{(1)}}{2}(\zeta+\delta\Delta)^{2}\notag\\
&+d_{L+}^{(0)}(\zeta-\delta\Delta)+\frac{d_{L+}^{(1)}}{2}(\zeta-\delta\Delta)^{2}+\mathcal{O}(\delta\Delta^{3}),\label{deldel_n_ss}\\
\frac{2}{g_{L}}\delta\Delta=&d_{L-}^{(0)}(\zeta+\delta\Delta)+\frac{d_{L-}^{(1)}}{2}(\zeta+\delta\Delta)^{2}\notag\\
&-d_{L+}^{(0)}(\zeta-\delta\Delta)-\frac{d_{L+}^{(1)}}{2}(\zeta-\delta\Delta)^{2}+\mathcal{O}(\delta\Delta^{3}).\label{deldel_d_ss}
\end{align}
By solving eqs. (\ref{deldel_n_ss}) and (\ref{deldel_d_ss}), we obtain $\zeta$ and $\delta\Delta$ as functions of $\delta n$ as
\begin{align}
\zeta \simeq&\frac{\frac{1}{g_{L}}-\frac{d_{L-}^{(0)}+{d_{L+}^{(0)}}}{2}}{d_{L-}^{(0)}+{d_{L+}^{(0)}}}\frac{1}{r}\delta n + \frac{1}{2}\frac{1}{\left(d_{L-}^{(0)}+{d_{L+}^{(0)}}\right)^{3}}\frac{C_{\mu}}{r^{3}}\delta n^{2},\label{deltamu_deltan}\\
\delta\Delta \simeq& \frac{1}{2}\frac{d_{L-}^{(0)}-{d_{L+}^{(0)}}}{d_{L-}^{(0)}+{d_{L+}^{(0)}}}\frac{1}{r}\delta n + \frac{1}{2}\frac{1}{\left(d_{L-}^{(0)}+{d_{L+}^{(0)}}\right)^{3}}\frac{C_{\Delta}}{r^{3}}\delta n^{2},\label{deltaDelta_deltan}
\end{align}
where the coefficients are defined as $r\equiv \frac{1}{g_{L}}-\frac{2d_{L-}^{(0)}{d_{L+}^{(0)}}}{d_{L-}^{(0)}+{d_{L+}^{(0)}}}$, $C_{\mu}\equiv d_{L-}^{(1)}\left(\frac{1}{g_{L}}-{d_{L+}^{(0)}}\right)^{3}+d_{L+}^{(1)}\left(\frac{1}{g_{L}}-d_{L-}^{(0)}\right)^{3}$ and $C_{\Delta}\equiv -d_{L-}^{(1)}{d_{L+}^{(0)}}\left(\frac{1}{g_{L}}-{d_{L+}^{(0)}}\right)^{2}+d_{L+}^{(1)}d_{L-}^{(0)}\left(\frac{1}{g_{L}}-d_{L-}^{(0)}\right)^{2}$. Here, we assume $r\neq 0$.
\par
By substituting (\ref{deltamu_deltan}) and (\ref{deltaDelta_deltan}) into (\ref{EEpv}), we obtain the expansion of $\mathcal{E}_{\eta}$ with respect to $\widetilde{\delta n}$ defined in eq. (\ref{widetilde_n}) as
\begin{eqnarray}
\mathcal{E}_{\eta}-\frac{g_{L}n_{L}}{2}\delta n\simeq {\rm const.}
+\widetilde{b_{\eta}}\widetilde{\delta n}^{2}+\widetilde{c_{\eta}}\widetilde{\delta n}^{3},
\end{eqnarray}
where $\eta=p,m$ and the coefficients are defined as
\begin{align}
&\widetilde{b_{m}}=\frac{1}{2}\frac{g}{d_{L-}^{(0)}+{d_{L+}^{(0)}}}\frac{\frac{1}{g^{2}}-d_{L-}^{(0)}{d_{L+}^{(0)}}}{\frac{1}{g}-\frac{2d_{L-}^{(0)}{d_{L+}^{(0)}}}{d_{L-}^{(0)}+{d_{L+}^{(0)}}}},\\
&\widetilde{b_{p}}=\frac{1}{2}\frac{1}{d_{L-}^{(0)}},\\
&\widetilde{c_{m}}=\frac{1}{6}\frac{1}{(d_{L-}^{(0)}+{d_{L+}^{(0)}})^{3}}\frac{1}{r^{3}}\nonumber\\
&\times\left[
\left\{
2\left(\frac{1}{g_{L}}-{d_{L+}^{(0)}}\right)^{3}+3\left(\frac{1}{g_{L}}-d_{L-}^{(0)}\right)\left(\frac{1}{g_{L}}-{d_{L+}^{(0)}}\right)^{2}
\right\}d_{L-}^{(1)}\right.\nonumber\\
+&\left.\left\{2\left(\frac{1}{g_{L}}-d_{L-}^{(0)}\right)^{3}+3\left(\frac{1}{g_{L}}-{d_{L+}^{(0)}}\right)\left(\frac{1}{g_{L}}-d_{L-}^{(0)}\right)^{2}
\right\}{d_{L+}^{(1)}}
\right],\\
&\widetilde{c_{p}}=\frac{5}{6}\frac{1}{{d_{L-}^{(0)}}^{3}}\left(1-\frac{3}{5}g_{L}d_{L-}^{(0)}\right)d_{L-}^{(1)}.
\end{align}
Then the uniform charge compressibility $\kappa$ around Lifshitz critical points is given as
\begin{eqnarray}
\kappa^{-1} &=& \left.\frac{\partial^{2}\widetilde{\mathcal{E}}_{\eta}}{\partial n^{2}}\right|_{n=n_{L}+\widetilde{\delta n}}\nonumber\\
&\simeq&2\widetilde{b_{\eta}}+6\widetilde{c_{\eta}}\widetilde{\delta n}.
\label{last_kappa}
\end{eqnarray}


\begin{thebibliography}{99}
\bibitem{note}In the present paper, we assume the uniqueness of the decomposition of the density of states into the regular part and the singular part, as eqs. (13) and (A.6). However, for the neck-collapsing case, we also obtain the unique expansion of the free energy on the more weak assumption. Namely $\forall |x|\ll 1, \exists \rho>0, \exists C>0$, if 
$ 
\rho\ln\frac{1}{|x|}+C_{1}\leq D_{-}(x) \leq \rho\ln\frac{1}{|x|}+C_{2}
$ is satisfied for $|C_{1}|<C$ and $|C_{2}|<C$, then, the coefficients $a$ and $b_{p}=b_{m}$ are uniquely determined by the densities of states around Lifshitz critical points.
\end{thebibliography}
\end{document}